\documentclass[acmsmall]{acmart}
\AtBeginDocument{%
  \providecommand\BibTeX{{%
    \normalfont B\kern-0.5em{\scshape i\kern-0.25em b}\kern-0.8em\TeX}}}

\setcopyright{acmcopyright}
\copyrightyear{2021}
\acmYear{2021}
\acmDOI{10.nn/nnnnnnn.nnnnnnn}

\acmJournal{JACM}
\acmVolume{1}
\acmNumber{1}
\acmArticle{111}
\acmMonth{1}

\usepackage{booktabs}
\usepackage{array}
\usepackage{balance} 
\usepackage{lipsum}
\usepackage{multirow}
\usepackage{booktabs}
\usepackage{amsmath}
\usepackage{subfigure}
\usepackage{algorithm}  
\usepackage{algorithmicx}  
\usepackage{algpseudocode}  
\usepackage{amsmath}  
\usepackage{enumitem}
\usepackage{tabularx}
\usepackage[utf8]{inputenc}
\usepackage[english]{babel}
\usepackage{amsthm}
\usepackage{bm}
\newcommand{\ie}{\emph{i.e., }}
\newcommand{\eg}{\emph{e.g., }}
\newcommand{\etal}{\emph{et al. }}

\newcommand{\wrt}{\emph{w.r.t. }}

\newlength\myindent
\usepackage[normalem]{ulem}
\useunder{\uline}{\ul}{}

\floatname{algorithm}{Algorithm}

\begin{document}

\title{Causal Disentangled Recommendation Against User Preference Shifts}


\author{Wenjie Wang}
\email{wenjiewang96@gmail.com}
\affiliation{%
  \institution{National University of Singapore}
  \country{Singapore}
}
\author{Xinyu Lin}
\email{xylin1028@gmail.com}
\affiliation{%
  \institution{National University of Singapore}
  \country{Singapore}
}
\author{Liuhui Wang}
\email{wangliuhui0401@pku.edu.cn}
\affiliation{%
  \institution{Peking University}
  \country{China}
}
\author{Fuli Feng}
\authornote{Corresponding author: Fuli Feng (fulifeng93@gmail.com).}
\email{fulifeng93@gmail.com}
\affiliation{%
  \institution{University of Science and Technology of China}
  \country{China}
}

\author{Yunshan Ma}
\email{yunshan.ma@u.nus.edu}
\affiliation{%
  \institution{National University of Singapore}
  \country{Singapore}
}

\author{Tat-Seng Chua}
\email{dcscts@nus.edu.sg}
\affiliation{%
  \institution{National University of Singapore}
  \country{Singapore}
}


\begin{abstract}

Recommender systems easily face the issue of user preference shifts. User representations will become out-of-date and lead to inappropriate recommendations if user preference has shifted over time. To solve the issue, existing work focuses on learning robust representations or predicting the shifting pattern. There lacks a comprehensive view to discover the underlying reasons for user preference shifts. To understand the preference shift, we abstract a causal graph to describe the generation procedure of user interaction sequences. Assuming user preference is stable within a short period, we abstract the interaction sequence as a set of chronological environments. From the causal graph, we find that the changes of some unobserved factors (\eg becoming pregnant) cause preference shifts between environments. Besides, the fine-grained user preference over categories sparsely affects the interactions with different items. Inspired by the causal graph, our key considerations to handle preference shifts lie in modeling the interaction generation procedure by: 1) capturing the preference shifts across environments for accurate preference prediction, and 2) disentangling the sparse influence from user preference to interactions for accurate effect estimation of preference. To this end, we propose a Causal Disentangled Recommendation (CDR) framework, which captures preference shifts via a temporal variational autoencoder and learns the sparse influence from multiple environments. Specifically, an encoder is adopted to infer the unobserved factors from user interactions while a decoder is to model the interaction generation process. Besides, we introduce two learnable matrices to disentangle the sparse influence from user preference to interactions. Lastly, we devise a multi-objective loss to optimize CDR. Extensive experiments on three datasets show the superiority of CDR in enhancing the generalization ability under user preference shifts. 

\end{abstract}



\maketitle

\section{Introduction}
\label{sec:introduction}

\begin{figure}[t]
\setlength{\abovecaptionskip}{0.2cm}
\setlength{\belowcaptionskip}{0.1cm}
\centering
\includegraphics[scale=0.45]{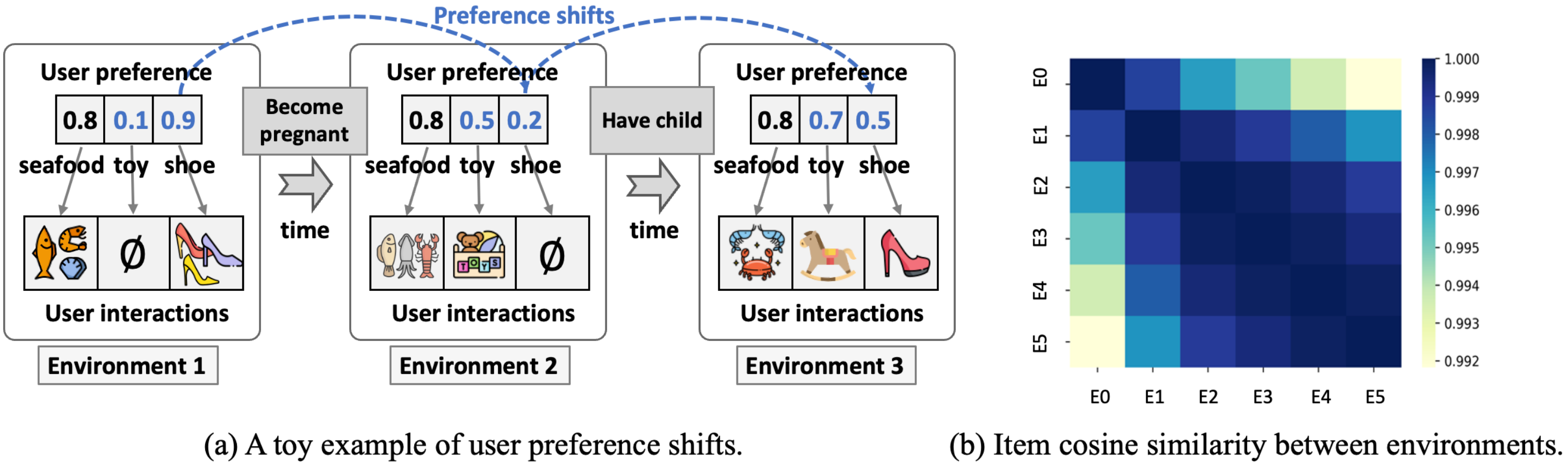}
  \caption{{(a) is a toy example to illustrate preference shifts. We assume that user preference is stable within a short period, thus treating a short time period as an environment. Some user preference is stable across environments, \eg the preference over seafood; while partial preference is shifting due to the changes of user features (\eg becoming pregnant).} {(b) shows the cosine similarity of item categories interacted by users in different environments of Amazon Book. The environments are chronologically split, starting from E0 to E5. It shows that item similarity decreases over time, revealing the interaction distribution shifts.}}
  \label{fig:example}
\end{figure}

Recommender models typically learn user preference representations from historical interactions (\eg clicks and ratings)~\cite{he2020lightgcn}. 
However, most recommender models assume the training and testing interactions are Independent and Identically Distributed (IID),
which is infeasible in real-world applications. 
As shown in Figure~\ref{fig:example}, there are user preference shifts across different environments~\cite{zafari2019modelling, wang2022causal} where each environment denotes a short time period. Some user features and environmental factors will change over time such as becoming pregnant, causing the shifts of user preference and interaction distributions in Out-of-Distribution (OOD) environments.
Such drifts can frustrate the recommender models trained over historical interactions~\cite{wang2022causal}. 
Consequently, the inferior performance of recommender models will degrade user experience and reduce user activities, hurting the health of the whole recommender system. 
Therefore, it is essential to capture user preference shifts and pursue generalizable recommendation.


Existing work mainly handles preference shifts from two perspectives: 
\begin{itemize}[leftmargin=*]
    \item \textit{Robust models against preference shifts.} The most representative approach is disentangled recommendation~\cite{ma2019learning, ma2020disentangled}, which disentangles several independent representations to represent different user preference. Disentangled representations are less sensitive to preference shifts~\cite{ma2019learning} since only partial representations are shifted while most are robust in the OOD environment. Nevertheless, existing work usually ignores the temporal shifting patterns of user preference across environments, limiting the generalization ability of recommender models. 
    
    \item \textit{Sequential models to predict the shifts.} Sequential recommendation~\cite{zhang2021cause, xie2021adversarial} recognizes the preference shifts by modeling the temporal patterns within the interaction sequence. 
    However, these sequential models typically overlook the sparse influence of user preference on interactions: partial preference shifts between environments only affect a small portion of user interactions. Ignoring such sparse influence may harm the interaction predictions over extensive irrelevant items, resulting in many inappropriate recommendations in OOD environments. 
\end{itemize}


There lacks a comprehensive view to reveal the underlying factors regarding user preference shifts. 
As such, we resort to causal language to inspect the causal relations behind the generation procedure of user interaction sequences. 
As shown in Figure~\ref{fig:causal_graph}, the interaction sequence is divided into multiple short time periods, where each period is viewed as an environment. 
$E_t$ represents the unobserved user features (\eg pregnancy) or environmental factors\footnote{Recommender models can observe few factors affecting preference due to privacy restriction and technical challenges~\cite{wang2018toward}.} (\eg hot events) in the environment $t$, which affect the user preference $Z_t$ and interactions $X_t$. 
Within each environment, user preference is relatively stable; however, some changes from $E_{t-1}$ to $E_t$ (\eg becoming pregnant) will shift user preference from $Z_{t-1}$ to $Z_t$. 
As to the effect of $Z_t$ on $X_t$, various category-level preference in $Z_t$ sparsely affects the interactions in $X_t$ as shown in the right part of Figure \ref{fig:causal_graph}. For instance, the preference over the category ``Toy'' influences the interactions with toy products. Due to the sparse influence, partial preference shifts from $Z_{t-1}$ to $Z_t$ only affect some interactions.

According to the causal relations, the key of handling user preference shifts lies in simultaneously 1) capturing the temporal shifts across environments ($Z_{t-1} \rightarrow Z_t$), \ie accurate preference prediction, and 2) disentangling the sparse influence from user preference to the interactions ($Z_t \rightarrow X_t$), \ie accurate effect estimation of preference. Disentangling such influence is essentially discovering the causal structure from  $Z_t$ to $X_t$~\cite{pearl2009causality, he2021daring}.
However, it is non-trivial to model the temporal preference shifts and the sparse influence due to the following challenges: 1) the changes of $E_t$ between environments are usually unobserved, hindering the accurate preference prediction; and 2) the causal structure from hidden user preference to interactions also lacks supervision, which requires us to find additional signals for the structure discovery between $Z_t$ and $X_t$.

To this end, we propose a Causal Disentangled Recommendation (CDR) framework, which models the interaction generation procedure in Figure~\ref{fig:causal_graph}. 
Specifically, 1) to estimate the unobserved $E_t$, CDR introduces a temporal Variational AutoEncoder (VAE), where an encoder uses variational inference to infer unobserved $E_t$ from observed interactions $X_t$. Besides, a decoder is to estimate the effect of $E_t$ on $X_t$ via $Z_t$, where $Z_t$ is iteratively updated to model the temporal shifts across environments. 
2) Furthermore, we introduce two learnable matrices to formulate the causal structure from $Z_t$ to $X_t$. The two matrices disentangle the representations of $Z_t$ into category-level preference, which then sparsely affects the interactions in corresponding categories. 
Due to lacking supervision, we propose to learn the two matrices from multiple environments, where the distribution shifts shed light on the sparse structure learning between $Z_t$ and $X_t$~\cite{scholkopf2021toward, liu2021heterogeneous}.
In particular, we utilize the variance regularization to balance the predictions across environments and adopt the sparsity regularization to control the sparsity of the structure.
Note that the VAE ignores the temporal information and fairly considers each interaction within each environment, and thus it captures the invariant preference in a short period. As such, during training and inference, we can flexibly adjust the division of environments to balance the modeling of invariant preference within environments and shifted preference between environments. 
Extensive experiments on three real-world datasets validate the effectiveness of CDR in capturing preference shifts and achieving superior performance in the OOD environments. We release the code and data at \url{https://github.com/Linxyhaha/CDR}.

The main contributions of this work are threefold:
\begin{itemize}[leftmargin=*]
    \item We retrospect user preference shifts across multiple environments from a causal view and inspect the underlying causal relations via a causal graph. 
    \item 
    We propose a CDR framework, which captures the preference shifts between environments via a temporal VAE and learns a sparse structure between user preference and interactions for the robust interaction prediction. 
    \item Empirical results on three public datasets demonstrate the superiority of CDR over the baselines \wrt the OOD generalization ability under preference shifts. 
    
\end{itemize}

\section{Method}
In this section, we present the causal mechanism and task formulation of recommendation with consideration of preference shifts in Section \ref{sec:problem}. 
Thereafter, we detail the proposed CDR framework in Section \ref{sec:method}. 

\subsection{Recommendation with Preference Shifts}
\label{sec:problem}

\begin{figure}[t]
\setlength{\abovecaptionskip}{0.2cm}
\setlength{\belowcaptionskip}{0cm}
\centering
\includegraphics[scale=0.7]{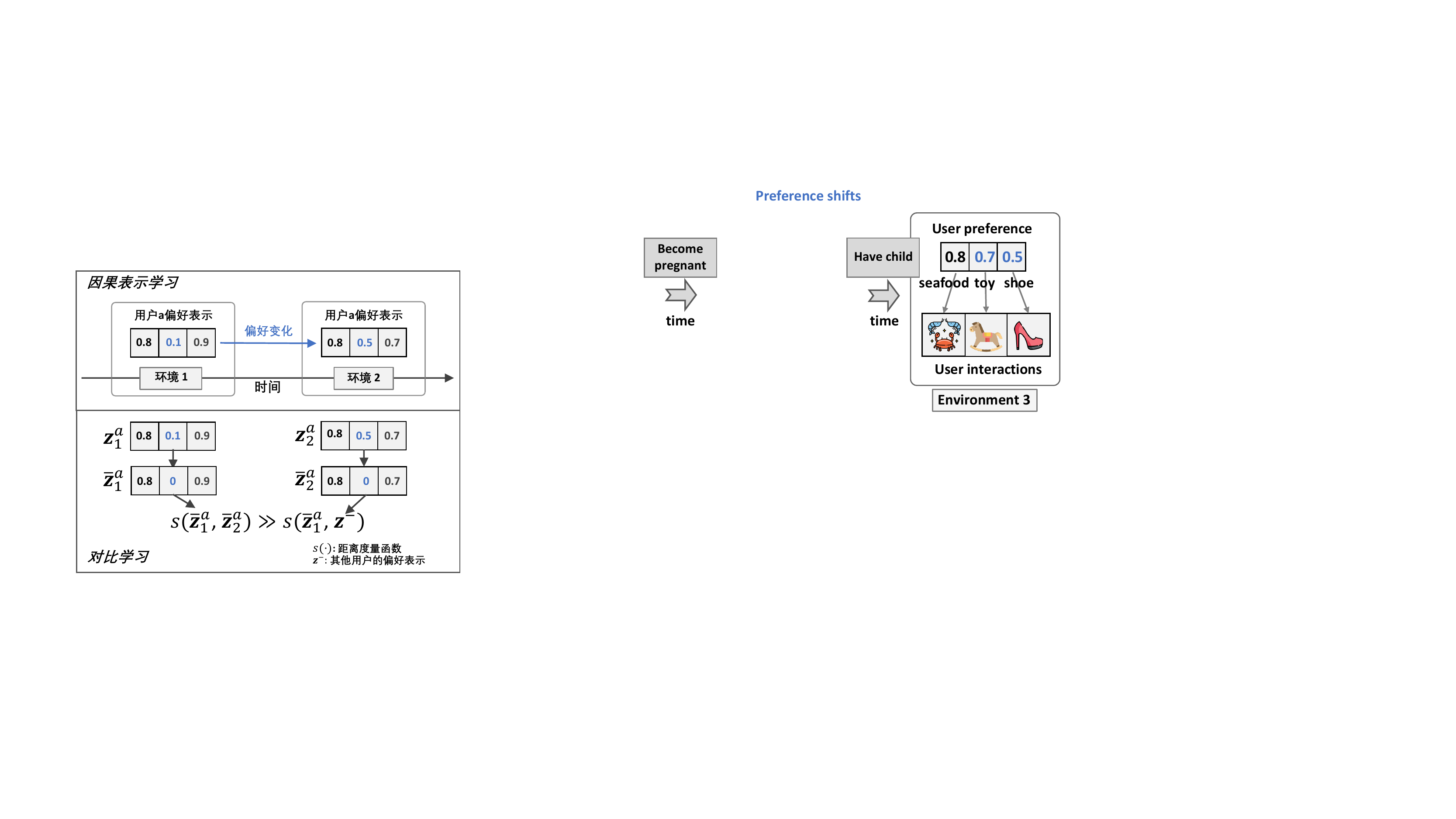}
\caption{{Causal graph behind the interaction generation procedure with multiple environments. We assume that the observed user preference is affected by some hidden user features, and various user preference sparsely controls the interactions with different items.}}
\label{fig:causal_graph}
\end{figure}

Existing recommender models usually rely on the IID interaction distributions from training to testing stages. Without considering user preference shifts over time, 
these recommender models will encounter significant performance drop in OOD environments. 
To improve the generalization ability, we build recommender models with considering the preference shifts. 
In this subsection, we first scrutinize the causal relations regarding user preference shifts, and then formulate the task of generalizable recommendation to evaluate the generalization ability under preference shifts. 

\subsubsection{\textbf{Causal View of Preference Shifts}}
We present the causal relations in Figure \ref{fig:causal_graph} and explain its rationality as follows:
\begin{itemize}[leftmargin=*]
    \item $E_t$ denotes unobserved user features (\eg pregnancy) or environmental factors (\eg hot events) in the environment $t$; $Z_t$ and $X_t$ represent the user preference and interactions, respectively. Because of the privacy restriction~\cite{wang2018toward}, we seldom utilize user features for recommendation, and thus we ignore the modeling of observed user features in Figure \ref{fig:causal_graph}, which can be easily incorporated as the input of the CDR framework if necessary. 
    
    \item $E_t \rightarrow Z_t$: user features and various environmental factors affect user preference.
    
    \item $Z_t \rightarrow X_t$: user interactions are determined by current user preference. {In particular, $Z_t$ covers the preference over multiple item categories (\eg seafood and toy). Some factors in $Z_t$ may represent the preference over an item category (\eg seafood), which sparsely affects a category of interactions as shown in Figure \ref{fig:causal_graph}}.
    
    \item $Z_{t-1} \rightarrow Z_t$: the user preference $Z_t$ in the environment $t$ is updated from previous $Z_{t-1}$, which exhibits the preference shifts over time. {From the causal graph, we find that various factors in $E_t$ can affect user preference $Z_t$ and cause the preference shifts $Z_{t-1} \rightarrow Z_t$, leading to the variation of user interaction distributions.
    Besides, the preference shifts between environments only influence partial interactions due to the sparse influence from $Z_t$ to $X_t$.}
    {\item $E_{t-1} \dashrightarrow E_t$ and $X_{t-1} \dashrightarrow Z_t$: $E_{t-1}$ might affect $E_{t}$ because user features might have conditional relations, \eg pregnancy $\rightarrow$ having child. Besides, user preference $Z_t$ can be influenced by previous interactions $X_{t-1}$. However, these conditional relations are not easy to be inferred from pure interactions, and the effects of these conditional relations on $Z_t$ and $X_t$ are relatively weaker than $(E_t, Z_{t-1})\rightarrow Z_t$ and $Z_t \rightarrow X_t$. As such, we omit the modeling of $E_{t-1} \dashrightarrow E_t$ and $X_{t-1} \dashrightarrow Z_t$ in this work to pursue a simple model with fewer parameters. Empirical evidence in Section~\ref{sec:condition_rel} also validates the superiority of our choice.}
\end{itemize}

\subsubsection{\textbf{Task Formulation}}\label{sec:task}
To evaluate the generalization ability under preference shifts, we formulate the task of generalizable recommendation. Formally, we utilize $u \in \{1,2,...,U\}$, $i\in \{1,2,...,I\}$, and $t\in \{1,2,...,T\}$ to index the user, item, and environment, respectively. The interactions of user $u$ in $T$ environments are denoted as $\bm{x}_{1:T}$\footnote{For notation brevity, we omit the subscript $u$ for $\bm{x}_{1:T}$ and $\bm{z}_{1:T}$ of user $u$.}, where $\bm{x}_{t}\in \{0,1\}^I$ is a multi-hot vector, and $x_{t,i}$ implies that user $u$ likes item $i$ ($x_{t,i}=1$) or not ($x_{t,i}=0$). 
Generally, given the observed $\bm{x}_{1:T}$ of user $u$, \textbf{generalizable recommendation} aims to capture the hidden preference shifts in $\bm{z}_{1:T}$ and estimates the latest user preference $\bm{z}_T$. 


\vspace{3pt}
\noindent$\bullet\quad$\textbf{Environment division.}
We can divide the environments by time, for instance, equally splitting the user interaction sequence into $T$ pieces, or clustering adjacent interactions according to the time interval. In this work, we choose the first one to simplify the data pre-processing.

\vspace{3pt}
\noindent$\bullet\quad$\textbf{{Inference for future environments.}}
{To evaluate the generalization ability of recommender models, we can utilize the interactions in the environment $T+1$ for testing. To infer the interaction probability in this unknown environment, we consider three strategies:
1) using the latest user preference $\bm{z}_T$ for prediction; 2) uniformly averaging the predictions in $T$ training environments; 3) considering the average user features $\bm{e}_{T+1}=\frac{1}{T}\sum^T_{t=1}\bm{e}_t$ and $\bm{z}_T$ to predict $\bm{z}_{T+1}$, and then using $\bm{z}_{T+1}$ for interaction prediction. Because the testing environment is unknown, these inference strategies inevitably make some assumptions. 
The first strategy requires the small preference shifts from environment $T$ to $T+1$. Meanwhile, the second and third strategies need the average over $T$ training environments, losing some temporal shifting patterns. In practice, we set the first strategy as the default due to its better performance on real-world datasets (refer to Section~\ref{sec:infer_unk}).
}

\vspace{3pt}
\noindent$\bullet\quad$\textbf{Difference from sequential recommendation.}
The main difference between generalizable and sequential recommendations is that generalizable recommendation emphasizes the preference shifts across environments and the invariant preference within an environment. Moreover, generalizable recommendation focuses on the predictions of multiple interactions in an OOD environment, which differs from the next-item prediction in sequential recommendation. 

\subsection{CDR Framework}
\label{sec:method}

In this subsection, we present the CDR framework to model the interaction generation procedure under multiple environments. In particular, we utilize a novel temporal VAE to capture the preference shifts ($Z_{t-1}\rightarrow Z_{t}$) and conduct sparse structure learning to disentangle the sparse influence from user preference to interactions ($Z_{t}\rightarrow X_t$). 

We construct the recommender model by following the causal relations in Figure \ref{fig:causal_graph}. Specifically, for each user $u$ in the environment $t$, we first sample a $K$-dimensional latent representation $\bm{e}_t$ from the standard Gaussian prior $\mathcal{N}\left(0, \mathbf{I}_{K}\right)$~\cite{liang2018variational, yang2021causalvae}, where the covariance $\mathbf{I}_{K}$ is an identity matrix. We then obtain user preference $\bm{z}_t \in \mathbb{R}^{H}$ based on $\bm{e}_t$ and the previous $\bm{z}_{t-1}$. Thereafter, $\bm{z}_t$ is used to predict the interaction probability over $I$ items and the historical interactions $\bm{x}_t\in \mathbb{R}^{I}$ are assumed to be drawn from the interaction probability distribution. In this work, we assume that $\bm{z}_t$ and $\bm{x}_t$ follow the factorized Gaussian and multinomial priors due to their superiority shown in previous work~\cite{liang2018variational, ma2019learning}.
Formally,
\begin{equation}
\label{eqn:all_prior}
\left\{
\begin{aligned}
& \bm{e}_t \sim \mathcal{N}\left(0, \mathbf{I}_{K}\right), \\
& \bm{z}_t \sim \mathcal{N}\left(\bm{\mu}_{\theta_1}(\bm{e}_t, \bm{z}_{t-1}), \text{diag}\{\bm{\sigma}^2_{\theta_1}(\bm{e}_t, \bm{z}_{t-1})\}\right), \\
& \bm{x}_t \sim \text{Mult}\left(N_t, \pi\left(f_{\theta_2}(\bm{z}_t)\right)\right). \\
\end{aligned}
\right.
\end{equation}
Specifically, we explain the generative process in Eq. (\ref{eqn:all_prior}), which is consistent with the causal relations in Figure \ref{fig:causal_graph}: 
\begin{itemize}[leftmargin=*]
    \item $(\bm{e}_t, \bm{z}_{t-1}) \rightarrow \bm{z}_t$: $\bm{\mu}_{\theta_1}(\bm{e}_t, \bm{z}_{t-1})$ and $\bm{\sigma}^2_{\theta_1}(\bm{e}_t, \bm{z}_{t-1})$ denote the \textit{mean} and \textit{diagonal covariance} of the Gaussian distribution of $\bm{z}_t$, which are estimated from $\bm{e}_t$ and $\bm{z}_{t-1}$ via a network $f_{\theta_1}(\cdot)$.
    \item $\bm{z}_t \rightarrow \bm{x}_t$: $\bm{x}_t$ is sampled from a multinomial distribution affected by $\bm{z}_t$, where $N_t=\sum_{i=1}^{I}x_{t,i}$ represents the interaction number of user $u$ in the environment $t$, $\pi(\cdot)$ is the \textit{softmax} function, and the network $f_{\theta_2}(\bm{z}_t)$ transforms $\bm{z}_t$ to produce the interaction probability over $I$ items. 
\end{itemize}

To train the recommender model, we aim to optimize the parameters $\{\theta_1, \theta_2\}$ by maximizing the generative probability of observed user interactions $\bm{x}_{1:T}$ in $T$ environments. {Formally, following~\cite{chung2015recurrent}, we can factorize the joint distribution $p(\bm{x}_{1:T})$ and maximize the log-likelihood as follows:} 
\begin{equation}
\label{eqn:log_likelihood}
\begin{aligned}
{\log p(\bm{x}_{1:T})} & = {\log \int p(\bm{x}_{1:T}|\bm{e}_{1:T})p(\bm{e}_{1:T})d\bm{e}_{1:T}} \\
                    &= {\log \int \prod_{t=1}^T p(\bm{x}_t|\bm{x}_{1:t-1},\bm{e}_{1:t})p(\bm{e}_{1:T})d\bm{e_{1:T}}},
\end{aligned}
\end{equation}
{where $p(\bm{x}_t|\bm{x}_{1:t-1},\bm{e}_{1:t})$ aligns with the generation procedure in Eq. (\ref{eqn:all_prior}) and we will further factorize it with $\bm{z}_{1:t}$ in the decoding process \ie Eq. (\ref{eqn:decoder}).
Nevertheless, maximizing Eq. (\ref{eqn:log_likelihood}) is intractable because it involves the integral over unobserved $\bm{e}_{1:T}$. 
To solve the problem, we embrace \textit{variational inference}~\cite{liang2018variational} to approximate $\log p(\bm{x}_{1:T})$ by using a variational distribution $q(\bm{e}_{1:T}|\cdot)$. Formally,}
\begin{subequations}
\label{eqn:ELBO}
\begin{align}
{\log p(\bm{x}_{1:T})} &= {\log \int \prod_{t=1}^T         p(\bm{x}_t|\bm{x}_{1:t-1},\bm{e}_{1:t})p(\bm{e}_{1:T})\frac{q(\bm{e}_{1:T}|\cdot)}{q(\bm{e}_{1:T}|\cdot)}d\bm{e_{1:T}}}\\
                    &\geq {\mathbb{E}_{q(\bm{e}_{1:T}|\cdot)} \left[\log \frac{\prod_{t=1}^Tp(\bm{x}_t|\bm{x}_{1:t-1},\bm{e}_{1:t})p(\bm{e}_{1:T})}{q(\bm{e}_{1:T}|\cdot)}\right] \quad (\text{ELBO})}\\ 
                    &= {\mathbb{E}_{q(\bm{e}_{1:T}|\cdot)}\left[\sum_{t=1}^T\left(\log p(\bm{x}_t|\bm{x}_{1:t-1},\bm{e}_{1:t})-\text{KL}\left[q(\bm{e}_t|\cdot)\|p(\bm{e}_{t})\right]\right)\right]},
\end{align}
\end{subequations}
where variational inference introduces the Evidence Lower BOund (ELBO) of Eq. (\ref{eqn:ELBO}a) by using $q(\bm{e}_{1:T}|\cdot)=\prod_{t=1}^Tq(\bm{e}_t|\cdot)$. Meanwhile, the first term in Eq. (\ref{eqn:ELBO}c) represents the probability of collecting observed $\bm{x}_t$ conditioned on $\bm{x}_{1:t-1}$ and $\bm{e}_{1:t}$ while the second term denotes the Kullback-Leibler (KL) divergence between the variational distribution $q(\bm{e}_t|\cdot)$ and the prior of $\bm{e}_t$. By maximizing the ELBO in Eq. (\ref{eqn:ELBO}c), we are able to increase the log-likelihood $\log p(\bm{x}_{1:T})$. 
Note that we avoid factorizing {$p(\bm{x}_{t}|\bm{x}_{1:t-1},\bm{e}_{1:t})$} with $\bm{z}_{1:t}$ in Eq. (\ref{eqn:ELBO}), and then we do not estimate the distribution of unobserved $\bm{z}_t$ and $\bm{e}_t$ simultaneously by variational inference. This is because we choose the alternative Monte Carlo (MC) sampling to efficiently approximate the posterior distribution of $\bm{z}_t$~\cite{chen2012monte}. {MC sampling constructs a random sampling of $\bm{z}_t$ (\ie draw samples from $p(\bm{z}_t|\bm{e}_t,\bm{z}_{t-1})$) to estimate its distribution, which avoids unnecessary prior hypothesis over the mean and covariance of $\bm{z}_t$ (see Eq. (\ref{eqn:decoder})).}

\begin{figure}[t]
\setlength{\abovecaptionskip}{0.2cm}
\setlength{\belowcaptionskip}{0cm}
\centering
\includegraphics[scale=0.6]{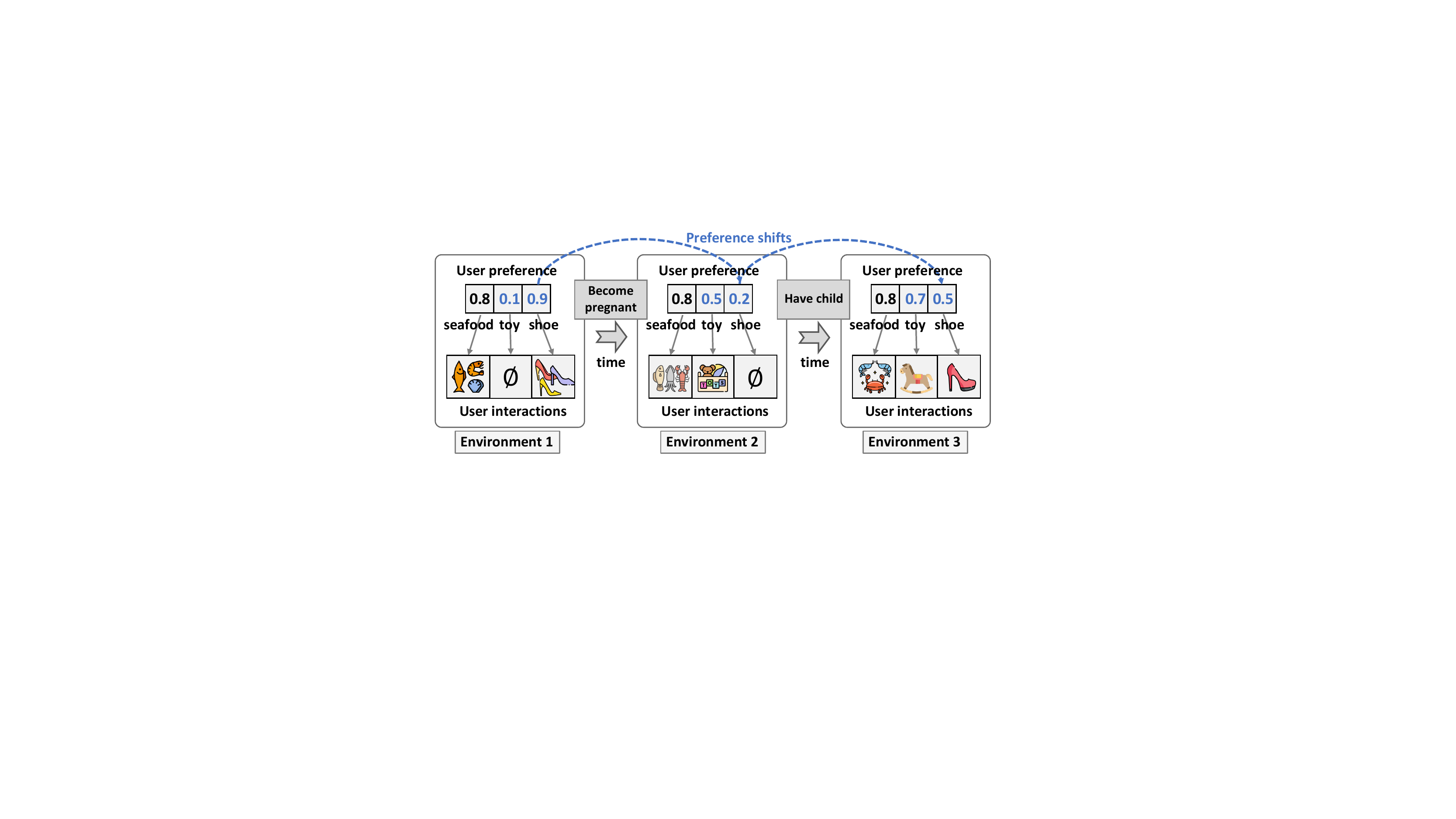}
\caption{Illustration of the CDR framework, where the encoder network predicts the hidden user features $\bm{e}_t$ in the environment $t$, and then the decoder reconstructs the interaction generation procedure from $\bm{e}_t$ and $\bm{z}_{t-1}$ to the interaction probability $f_{\theta_2}(\bm{z}_t)$. The entire encoder-decoder process repeats T times while $\bm{W}_z$ and $\bm{W}_x$ are shared across $T$ environments.} 
\label{fig:CDR}
\end{figure}

So far, the key of calculating the ELBO in Eq. (\ref{eqn:ELBO}c) lies in estimating $q(\bm{e}_t|\cdot)$ and {$p(\bm{x}_t|\bm{x}_{1:t-1},\bm{e}_{1:t})$}, which can be obtained by the encoder and decoder networks, respectively. We present the intuitive illustration of CDR with the encoder and decoder networks in Figure \ref{fig:CDR}.

\subsubsection{\textbf{Encoder Network}}

To estimate $q(\bm{e}_t|\cdot)$, we incorporate an encoder network $g_{\phi}(\cdot)$, which predicts $\bm{e}_t$ by the user interaction $\bm{x}_t$. The underlying motivation is that unobserved factors (\eg income) can be inferred from users' behaviors (\eg purchasing expensive products). In particular, 
\begin{equation}
\label{eqn:encoder}
\begin{aligned}
q(\bm{e}_t|\cdot) = q(\bm{e}_t|\bm{x}_t) = \mathcal{N}\left(\bm{e}_t; \bm{\mu}_{\phi}(\bm{x}_t), \text{diag}\{\bm{\sigma}^2_{\phi}(\bm{x}_t)\}\right),
\end{aligned}
\end{equation}
where $\bm{\mu}_{\phi}(\bm{x}_t)$ and $\bm{\sigma}^2_{\phi}(\bm{x}_t)$ denote the mean and diagonal covariance of $\bm{e}_t$, respectively. They are estimated by the encoder network $g_\phi(\cdot)$ parameterized by $\phi$. Formally, we have $g_{\phi}(\bm{x}_t) = [\bm{\mu}_{\phi}(\bm{x}_t), \bm{\sigma}_{\phi}(\bm{x}_t)]\in \mathbb{R}^{2K}$. In this work, we instantiate $g_\phi(\cdot)$ by a Multi-Layer Perceptron (MLP), which outputs the Gaussian parameters of $\bm{e}_t$. Note that the encoder ignores the temporal interaction sequence in $\bm{x}_t$ and fairly encodes every interaction.

\subsubsection{\textbf{Decoder Network}}
We factorize {$p(\bm{x}_t|\bm{x}_{1:t-1},\bm{e}_{1:t})$} by following the causal relations in the interaction generation process:
\begin{equation}
\label{eqn:decoder}
\begin{aligned}
{p(\bm{x}_t|\bm{x}_{1:t-1},\bm{e}_{1:t})} &= {\int p(\bm{x}_t|\bm{z}_t)\prod_{a=1}^tp(\bm{z}_a|\bm{z}_{a-1},\bm{e}_a)d\bm{z}_{1:t}},
\end{aligned}
\end{equation}
{where $p(\bm{z}_a|\bm{z}_{a-1},\bm{e}_a)$ denotes the probability distribution of the user preference $\bm{z}_a$ in the environment $a$; and when $a=1$, $\bm{z}_{0}$ is set as the constant vector $\bm{0}$. Besides, to approximate the distribution of $\bm{z}_{t}$, we use MC sampling~\cite{chen2012monte} to draw samples from $p(\bm{z}_t|\bm{z}_{t-1}, \bm{e}_t)$. Then we can calculate $p(\bm{x}_t|\bm{x}_{1:t-1},\bm{e}_{1:t})$ based on $p(\bm{x}_t|\bm{z}_t)$ while marginalizing over $\bm{z}_{1:t}$ via the samples from MC sampling. 
To iteratively calculate $p(\bm{z}_t|\bm{z}_{t-1}, \bm{e}_t)$, we adopt an MLP model $f_{\theta_1}(\cdot)$ to output the $\bm{\mu}_{\theta_1}(\cdot)$ and $\bm{\sigma}_{\theta_1}(\cdot)$ of $\bm{z}_t$. Formally, we have $f_{\theta_1}(\bm{z}_{t-1}, \bm{e}_t) = \left[\bm{\mu}_{\theta_1}(\bm{z}_{t-1}, \bm{e}_t), \bm{\sigma}_{\theta_1}(\bm{z}_{t-1}, \bm{e}_t)\right] \in \mathbb{R}^{2H}$. Thereafter, the remaining challenge is estimating $p(\bm{x}_t|\bm{z}_t)$ in Eq. (\ref{eqn:decoder}).}

\vspace{10pt}
\noindent$\bullet\quad$\textbf{Sparse structure learning.} 
To estimate $p(\bm{x}_t|\bm{z}_t)$, we incorporate $f_{\theta_2}(\cdot)$ to transform $\bm{z}_t$ into the interaction probability over $I$ items. However, to align with the causal relations in Figure \ref{fig:causal_graph} and learn the sparse influence from user preference to interactions, we do not simply use an MLP model for the implementation of $f_{\theta_2}(\cdot)$. We instead resort to \textit{sparse structure learning} in multiple environments, which aims to discover a sparse structure from user preference representations to interactions and requires the structure is robust across all the environments with distribution shifts. Consequently, 1) the sparse structure learned from multiple environments instead of one environment will encode the robust relations between user representations and interactions, which are likely to be reliable in future environments with preference shifts; and 2) if partial user preference has shifted, only a subset of interactions are affected due to the sparse structure. Such characteristics will improve the generalization ability of CDR under preference shifts. 

\begin{figure}[t]
\setlength{\abovecaptionskip}{0.2cm}
\centering
\includegraphics[scale=0.8]{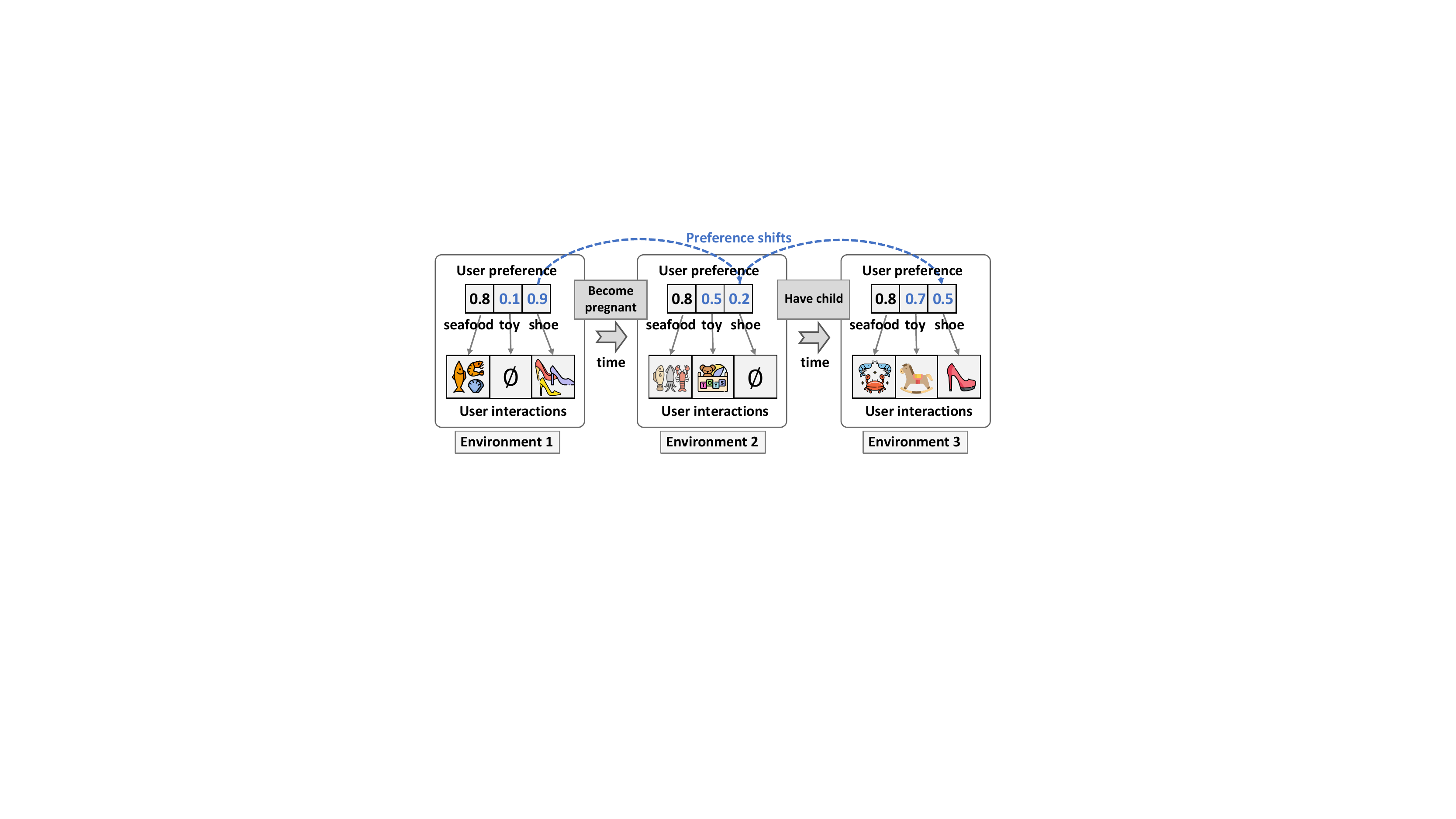}
\caption{Illustration of the calculation of $f_{\theta_2}(\bm{z}_t)$ in Eq. (\ref{eqn:f_theta_2}). Similar to masking mechanisms, $\bm{W}_z$ and $\bm{W}_x$ disentangle the user preference representations, leading to sparse connection from user preference to interactions. Note that we simplify $\bm{W}_z$ and $\bm{W}_x$ as discrete matrices with $\{0,1\}$ for better understanding.}
\label{fig:theta_2}
\end{figure}

Following~\cite{ma2019learning}, {the user preference representation $\bm{z}_t\in \mathbb{R}^{H}$ can cover the preference over multiple item categories, and we disentangle  $\bm{z}_t$ into several category-specific preference representations.}
Specifically, to implement the sparse structure, we introduce a matrix $\bm{W}_z \in \mathbb{R}^{H\times C}$ to factorize the user representation into the preference over $C$ item categories. In particular, ${W}_z[h,c]\in \bm{W}_z$ denotes the probability of the $h$-th factor in $\bm{z}_t$ belonging to the preference over the $c$-th category. 
Correspondingly, we leverage a matrix $\bm{W}_x \in \mathbb{R}^{I\times C}$ to classify items into $C$ categories. Inspired by~\cite{yamada2020feature, liu2021heterogeneous}, we draw $\bm{W}_z$ and $\bm{W}_x$ from the clipped Gaussian distributions parameterized by $\bm{\alpha}\in \mathbb{R}^{H\times C}$ and $\bm{\beta}\in \mathbb{R}^{I\times C}$, respectively. Formally, for each ${W}_z[h,c]\in \bm{W}_z$ and ${W}_x[i,c]\in \bm{W}_x$, we have 
\begin{equation}
\label{eqn:W_z_W_x}
\left\{
\begin{aligned}
{W}_z[h,c] &= \min\left(\max\left(\alpha_{h,c} + \epsilon, 0\right), 1\right), \\
{W}_x[i,c] &= \min\left(\max\left(\beta_{i,c} + \epsilon, 0\right), 1\right), \\
\end{aligned}
\right .
\end{equation}
where the noise $\epsilon$ is drawn from $\mathcal{N}\left(0, \sigma_{\epsilon}^2\right)$. We clip the values of $\bm{W}_z$ and $\bm{W}_x$ into $[0,1]$ to ensure a valid range for the probabilities. Besides, to encourage each factor or item belonging to one category, we add a softmax function at the dimension of $C$ categories in $\bm{W}_z$ and $\bm{W}_x$. 
As illustrated in Figure \ref{fig:theta_2}, we then implement $f_{\theta_2}(\cdot)$ to estimate the parameters of $\bm{x}_t$ in Eq. (\ref{eqn:all_prior}) by
\begin{equation}
\label{eqn:f_theta_2}
\begin{aligned}
f_{\theta_2}(\bm{z}_t) = \sum_{c=1}^C {W}_x[:,c] \odot f_{\gamma}({W}_z[:,c] \odot \bm{z}_t),
\end{aligned}
\end{equation}
where $\theta_2=\{\bm{\alpha}, \bm{\beta}, \gamma\}$, $\odot$ denotes the element-wise multiplication, and $f_{\gamma}(\cdot)$ can be any function transforming $\bm{z}_t$ to the interaction probability distribution over $I$ items. Following~\cite{liang2018variational}, we implement $f_{\gamma}(\cdot)$ by an MLP model.

\vspace{10pt}
\noindent$\bullet\quad$\textbf{Likelihood estimation.} 
As shown in Figure~\ref{fig:CDR}, given the user interactions $\bm{x}_{1:T}$ of user $u$, we feed them into the encoder network to sample $\bm{e}_{1:T}$, and then iteratively pass $\bm{e}_{1:T}$ to the decoder network to obtain the parameters of the multinomial distribution for $\bm{x}_{1:T}$ (\ie $f_{\theta_2}(\bm{z}_{1:T})$). Thereafter, {the log-likelihood $\log p(\bm{x}_t|\bm{z}_{t})$ can be calculated by} 
\begin{equation}
\label{eqn:likelihood}
\begin{aligned}
{\log p(\bm{x}_t|\bm{z}_{t})} &\overset{c}{=} \sum_{i=1}^{I}x_{t,i}\log \pi_i\left(f_{\theta_2}(\bm{z}_{t})\right),
\end{aligned}
\end{equation}
where $x_{t,i} \in \{0,1\}$ indicates whether user $u$ has interacted with item $i$ in the environment $t$ or not. 
Besides, the softmax function $\pi(\cdot)$ is used to normalize $f_{\theta_2}(\cdot)$ and $\pi_i(f_{\theta_2}(\cdot))$ denotes the normalized prediction score for item $i$. {Intuitively, the log-likelihood $p(\bm{x}_t|\bm{z}_{t})$ estimates the probability of drawing observed $\bm{x}_t$ from the multinomial distribution by sampling $N_t$ times, where $N_t$ is the interaction number of user $u$ in the environment $t$.}


\begin{algorithm}[t]
	\caption{Training of CDR under Multiple Environments}  
	\label{algo:training}
	\begin{algorithmic}[1]
		\Require $X_{1:T}$ of all $U$ users; $g_\phi(\cdot)$, $f_{\theta_1}(\cdot)$, and $f_{\theta_2}(\cdot)$ with initialized $\phi$, $\theta_1$, and $\theta_2$, respectively.
		\While{\textit{not converged}}
		    \State Sample a batch of users $\mathcal{U}$
		    \ForAll{$u \in \mathcal{U}$}
		        \ForAll{$t \in \{1,2,...,T\}$}
		            \State Sample $\bm{e}_t$ by feeding $\bm{x}_t$ into $g_\phi(\bm{x}_t)$;
    		        \State Sample $\bm{z}_t$ by feeding $\bm{z}_{t-1}$ and $\bm{e}_t$ into $f_{\theta_1}(\bm{z}_{t-1}, \bm{e}_t)$;
    		        \State Calculate $f_{\theta_2}(\bm{z}_t)$ via Eq. (\ref{eqn:f_theta_2});
    		        \State Obtain the probability of drawing $\bm{x}_t$ by Eq. (\ref{eqn:likelihood});
    		        \State Calculate the gradients \wrt the loss in Eq. (\ref{eqn:overall_loss});
                \EndFor
		    \EndFor
		    \State Average the gradients over $|\mathcal{U}|$ users and $T$ environments;
		    \State Update $\phi$, $\theta_1$, and $\theta_2$ via Adam;
		\EndWhile
		\Ensure $g_\phi(\cdot)$, $f_{\theta_1}(\cdot)$, and $f_{\theta_2}(\cdot)$.
	\end{algorithmic}
\end{algorithm}
\vspace{-0.1cm}

\subsubsection{\textbf{CDR Optimization}}
We maximize the ELBO to increase the log-likelihood in Eq. (\ref{eqn:log_likelihood}) by optimizing the parameters (\ie $\phi$ and $\theta=\{\theta_1, \theta_2\} $) in CDR. The parameters are updated by stochastic gradient descent. However, we conduct the sampling of $\bm{e}_t$ in Eq. (\ref{eqn:encoder}) and $\bm{z}_t$ in Eq. (\ref{eqn:decoder}), which prevents the back-propagation of gradients. To solve this problem, we utilize the \textit{reparameterization trick}~\cite{Kingma2014Auto, liang2018variational}. Besides, we leverage the \textit{KL annealing trick}~\cite{liang2018variational} to control the effect of the KL divergence, which introduces an additional hyper-parameter $\lambda_1$ into Eq. (\ref{eqn:ELBO}c). To summarize, the optimization objective for user $u$ is to minimize the following loss:
\begin{equation}
\label{eqn:rec_loss}
\notag
\small
\begin{aligned}
{\mathcal{L}^u} = {-\mathbb{E}_{q_{\phi}(\bm{e}_{1:T}|\cdot)} \left[\sum_{t=1}^T\left(\log p(\bm{x}_t|\bm{x}_{1:t-1},\bm{e}_{1:t})-\lambda_1\text{KL}\left[q(\bm{e}_t|\bm{x}_t)\|p(\bm{e}_t)\right]\right)\right]},
\end{aligned}
\end{equation}
{which becomes negative timestamp-wise ELBO~\cite{chung2015recurrent} over the $T$ environments.} 

\vspace{10pt}
\noindent$\bullet\quad$\textbf{Sparsity and variance regularization.}
In addition to the ELBO objective, we additionally consider two regularization terms: 1) the sparsity of $\bm{W}_z$ and $\bm{W}_x$, and 2) the variance of the gradients across $T$ environments. 
In Eq. (\ref{eqn:f_theta_2}), we expect that the structure implemented by $\bm{W}_z$ and $\bm{W}_x$ is sparse because the sparse connection between user representations and interactions is more robust under preference shifts. Therefore, we introduce an $L_0$ regularization term $\|\bm{W}_z\|_0 + \|\bm{W}_x\|_0$ to restrict the number of non-zero values. 

As to the variance regularization, it can facilitate the sparse structure learning across multiple environments. Training over multiple environments easily leads to imbalanced optimization: the performance in some environments is good while in other environments has inferior results. Consequently, disentangled preference representations via $\bm{W}_z$ and $\bm{W}_x$ might not be reliable across multiple environments. 
As such, we incorporate the variance penalty regularizer used in invariant learning~\cite{koyama2021invariance, liu2021heterogeneous}, which regulates the variance of the gradients under $T$ environments. Specifically, we calculate the variance regularization for user $u$ by $\sum_{t=1}^T\left\| \nabla_\theta \mathcal{L}^u_t - \nabla_\theta \mathcal{L}^u\right\|^2$,
where $\mathcal{L}^u_t$ is the optimization loss for the environment $t$ in Eq. (\ref{eqn:rec_loss}), $\nabla_\theta$ denotes the gradients \wrt the learnable parameters $\theta$, and $\nabla_\theta \mathcal{L}^u$ represents the average gradients over $T$ environments.

Intuitively, the variance regularizer will restrict the gradient difference among $T$ environments, and thus update the parameters $\theta$ synchronously for multiple environments. This will alleviate the problem that the parameters are unfairly optimized to improve the performance of few environments~\cite{liu2021heterogeneous}. To sum up, we have the final optimization loss for user $u$ as follows:
\begin{equation}
\label{eqn:overall_loss}
\begin{aligned}
\mathcal{L}^u + \lambda_2 \cdot (\|\bm{W}_z\|_0 + \|\bm{W}_x\|_0) + \lambda_3 \cdot \sum_{t=1}^T\left\| \nabla_\theta \mathcal{L}^u_t - \nabla_\theta \mathcal{L}^u\right\|^2,
\end{aligned}
\end{equation}
where two hyper-parameters $\lambda_2$ and $\lambda_3$ control the strength of sparsity and variance regularization terms, respectively. 



\subsubsection{\textbf{Environment division}}\label{sec:environment}
For the temporal interaction sequence of user $u$, we split it into $T$ pieces according to the equal interaction number in every environment. 
The choice of $T$ is essential because it balances the learning of shifted and invariant preference. CDR only considers the cross-environment preference shifts and assumes the intra-environment preference is invariant by ignoring the temporal information of interactions. 
Therefore, a larger $T$ will expose more sequential information to CDR. 
Nevertheless, the large $T$ value will increase the sparsity of interactions in each environment, hurting the learning of the encoder and decoder networks. To alleviate the dilemma, we choose a relatively small $T$ during training to ensure the interaction density of each environment. Once the parameters are well learned, we adopt a larger $T$ to fully utilize the sequential information in the inference stage.

\subsubsection{\textbf{Summary}}\label{sec:summary}
The detailed training procedure can be found in Algorithm~\ref{algo:training}.
To train the encoder and decoder networks in CDR, we divide the training interactions into multiple environments and utilize them to minimize the loss function in Eq. (\ref{eqn:overall_loss}) over all users. 
During the inference stage, we use the latest user preference $\bm{z}_T$ to calculate $f_{\theta_2}(\bm{z}_T)$ for the ranking of item candidates, and then recommend top-ranked items to each user.

To summarize, the encoder network infers unobserved $E_t$ from users' interactions. Thereafter, the decoder network leverages the inferred $E_t$ to iteratively update $Z_t$ for better preference estimation. Besides, the decoder network conducts sparse structure learning to model the sparse influence from $Z_t$ to $X_t$ for better effect estimation of user preference. As compared to traditional VAE-based methods~\cite{liang2018variational, wang2021personalized, xia2021Collaborative}, CDR is more robust in OOD environments because it constructs the encoder and decoder networks by following the causal relations in Figure \ref{fig:causal_graph}. 
Besides, thanks to modeling causal relations, CDR supports the intervention over the causal graph. {As illustrated in Section~\ref{sec:case_do_E}, we can estimate the counterfactual user preference $Z_t$ and the corresponding recommendations by intervening on $E_t$, \ie changing $E_t={\bm{e}}_t$ to $do(E_t=\hat{\bm{e}}_t)$~\cite{pearl2009causality}.} 

\section{Experiments}
\label{sec:experiment}
In this section, we conduct extensive experiments on three public datasets to answer the following research questions:
\begin{itemize}[leftmargin=*]
    \item \textbf{RQ1:} How does CDR perform under user preference shifts as compared to the baselines?
    \item {\textbf{RQ2:} How can the different designs in CDR (\eg the environment numbers, the sparse structure, inference strategies, multi-objective loss, and hyper-parameters) affect the performance?}
    \item {\textbf{RQ3:} How can we intuitively understand the effectiveness of CDR by case studies?}
\end{itemize}

\subsection{Experimental Settings}
\vspace{1pt}
\noindent$\bullet\quad$\textbf{Datasets.}
We evaluate the baselines and the proposed CDR on three real-world datasets: 1) Yelp\footnote{\url{https://www.yelp.com/dataset.}} is a public restaurant recommendation dataset, which contains rich interaction features such as ratings and timestamps; 2) Book is one of the Amazon product review datasets\footnote{\url{https://jmcauley.ucsd.edu/data/amazon/.}}, which covers extensive users' ratings over books;
and 3) Electronics is also from the Amazon datasets, in which users interact with various electrical products.  

The statistics of datasets are summarized in Table~\ref{tab:datasets_statics}.
To ensure the data quality~\cite{he2020lightgcn}, we only keep the users and items with at least 20 interactions on Yelp and Book. Besides, we only discard the users and items with less than 10 interactions on Electronics because the numbers of users and items are relatively small as shown in Table \ref{tab:datasets_statics}. Moreover, only the interactions with ratings $\geq 4$ are considered as positive samples on all three datasets. We sort user interactions chronologically, and then split the interactions of each user by the ratio of $80\%$, $10\%$, $10\%$ into training, validation, and test sets, respectively.

\begin{table}[t]
\setlength{\abovecaptionskip}{0cm}
\setlength{\belowcaptionskip}{0cm}
\caption{Statistics of the three datasets.}
\label{tab:datasets_statics}
\begin{center}
\resizebox{0.5\textwidth}{!}{
\begin{tabular}{lllll}
\toprule
{Dataset} & {\#User} & {\#Item} & {\#Interaction} & {Density} \\ \hline 
{Yelp} & 11,622 & 9,095 & 487,000 & 0.004607 \\ 
{Book} & 21,923 & 23,773 & 1,125,676 & 0.002159 \\
{Electronics} & 9,279 & 6,065 & 158,979 & 0.002825\\ \bottomrule
\end{tabular}
}
\end{center}
\vspace{-0.5cm}
\end{table}

\vspace{5pt}
\noindent$\bullet\quad$\textbf{Baselines.}
We compare CDR with the state-of-the-art collaborative filtering, disentangled and sequential models.

\textbf{- MF}~\cite{rendle2009bpr} is one of the most influential collaborative filtering methods, which factorizes the sparse interaction matrix into the user and item embedding matrices. 

\textbf{- LightGCN}~\cite{he2020lightgcn} is a powerful GCN-based recommender model, which discards the useless feature transformation and nonlinear activation in GCN, and highlights the most essential neighborhood aggregation for collaborative filtering.

\textbf{- MultiVAE}~\cite{li2017collaborative} is the most representative VAE-based recommender model, which captures the interaction generation process but ignores the temporal preference shifts.

\textbf{- MacridVAE}~\cite{ma2019learning} proposes the disentangled user representations at the intention and preference levels. The disentanglement enhances the model robustness against preference shifts. 

{\textbf{- DIB}~\cite{liu2021mitigating} utilizes information theory to disentangle biased and unbiased embeddings, and only considers unbiased embeddings for robust interaction prediction.}

{\textbf{- COR}~\cite{wang2022causal} proposes a causal OOD framework to handle the observed user feature shifts.}

{\textbf{- DIEN}~\cite{zhou2019deep} focuses on a new structure to model the interest evolving process, leading to more expressive user representations.}

{\textbf{- MGS}~\cite{lai2022attribute} applies a session graph generated from the user interaction sequence to capture transition patterns of user preference. We do not use user and item features in COR, DIEN, and MGS for fair comparison with other methods.}

\textbf{- DSSRec}~\cite{ma2020disentangled} introduces the techniques of self-supervised learning and disentangled representations to sequential recommendation. However, it ignores the advantages of learning disentangled representations from multiple environments.

\textbf{- ACVAE}~\cite{xie2021adversarial} is one of the state-of-the-art sequential recommender models, which incorporates contrastive learning and adversarial training into the VAE-based method. 

\textbf{- CauseRec}~\cite{zhang2021cause} constructs counterfactual sequences by keeping the indispensable interactions and replacing the dispensable ones. These counterfactual sequences are then used as augmented samples for contrastive training. 

We omit more sequential models such as GRU4Rec~\cite{GRU4Rec} and BERT4Rec~\cite{Bert4Rec} since ACVAE and CauseRec have shown better performance than them. 

\vspace{5pt}
\noindent$\bullet\quad$\textbf{Evaluation.}
We follow the all-ranking protocol~\cite{he2020lightgcn} to evaluate the performance of all methods, where all non-interacted items are used for ranking and top-ranked items are returned as recommendations. Thereafter, we adopt Recall@$K$ (R@$K$) and NDCG@$K$ (N@$K$) as the evaluation metrics, where $K = 10$ or $20$ on three datasets. 

\vspace{5pt}
\noindent$\bullet\quad$\textbf{Hyper-parameter settings.}
Based on the default settings of baselines, we enlarge their hyper-parameter search scope and tune hyper-parameters as follows:

{- MF}{ \& LightGCN}: The learning rate is searched in $\{0.001, 0.01, 0.1\}$. We search the best embedding size from $\{32, 64, 128\}$. For LightGCN, we tune the weight decay in $\{1e^{-5}, 1e^{-4}, 1e^{-3}\}$ and the number of GCN layers in $\{3, 4, 5\}$.

{- MultiVAE}{ \& MacridVAE}: We follow the default settings and additionally tune the learning rate, the hidden size, the regularization coefficient $\beta$ in $\{1e^{-4}, 1e^{-3}, 1e^{-2}\}$, $\{[800], [600, 200], [800, 500]\}$, and \{0.3, 0.5, 0.7, 0.9\}, respectively. As to special hyper-parameters in MacridVAE, we choose the number of macro factors from \{2, 4, 10, 20\}, the number of micro factors from \{200, 300, 400, 500\}, and the coefficient $\tau$ from \{0.05, 0.1, 0.2\}.

{{- DIB}{ \&COR}: For DIB, the embedding size is set in \{32, 64, 128\}. We adjust $\alpha$ in \{0.001, 0.1, 0\}, $\beta$ in \{0.0001, 0.001, 0.01, 0.1\} and $\gamma$ in \{0, 0.1, 0.2\} to make a balance between the biased and unbiased vector. For COR, we search the hidden size of encoder $q(\cdot)$ and the KL coefficient $\beta$ in $\{[300], [800], [800,600]\}$, and $\{0.3,0.5,0.7,0.9\}$, respectively. The sizes of $Z_1$ and $Z_2$ are chosen in $\{100,200,300\}$.}

{{- DIEN}{ \& MGS}: For DIEN, the embedding size is set in $\{32, 64, 128, 256, 512\}$. We search the dropout ratio in $\{0.1, 0.3, 0.5, 0.6\}$ and the weight of auxiliary loss in $\{0, 0.4, 0.6, 0.8, 1.0\}$. For MGS, we choose the embedding size from $\{50, 100, 200\}$, the number of GNN layers from $\{1, 3, 5, 6\}$, and the sequence length from $\{10, 30, 50\}$.}

{- DSSRec}{ \& ACVAE}{ \& CauseRec}: We search the training sequence length and the embedding size in $\{50, 100, 200\}$ and $\{32, 64, 128\}$, respectively. For DSSRec, the number of latent categories $K$ is chosen from $\{1, 2, 4, 8\}$ and the weight decay is set by $\{0, 0.01, 0.05\}$. For ACVAE, the weight of contrastive loss term $\beta$ is searched in $\{0.1, 0.3, 0.5, 0.7\}$. For CauseRec, the number of concepts and weight decay are tuned in $\{20, 30, 40\}$ and $\{1e^{-5}, 1e^{-4}, 1e^{-3}\}$, respectively.

As to CDR, we implement it by Pytorch and utilize Adam for optimization. For fair comparison, we choose the hyper-parameters by following the settings of the baselines. 
We set the batch size as $500$ and the learning rate as $1e^{-4}$. The dropout ratio is chosen from $\{0.4, 0.5, 0.6\}$. The hidden size of $g_{\phi}(\cdot)$ is searched in $\{[800], [600, 200], [800, 500]\}$. The sizes of $\bm{e}_t$ and $\bm{z}_t$ are tuned in $\{200, 300, 400, 500\}$. Both $f_{\theta_1}(\cdot)$ and $f_{\gamma}(\cdot)$ are set as a fully-connected layer to save parameters. $T$, $\lambda_1$, $\lambda_2$, $\lambda_3$ and $C$ are tuned in $\{1, 2, ..., 6\}$, $\{0.1, 0.2, ..., 0.9\}$, $\{0.1, 0.2, ..., 1\}$, $\{1e^{-5}, 1e^{-4}, 1e^{-3}, 1e^{-1}\}$, and $\{1, 2, 3, 4, 10, 20\}$, respectively. Moreover, early stopping is performed for model selection, \ie stop training if recall@10 on the validation set does not increase for 10 successive epochs. More details can be found in the released code.

\begin{table*}[t]
\setlength{\abovecaptionskip}{0.1cm}
\setlength{\belowcaptionskip}{0cm}
\caption{The performance comparison between the baselines and CDR on the three datasets. The best results are highlighted in bold and the second-best ones are underlined. \%improve. indicates the relative improvements of CDR than the second-best results. $*$ implies the improvements over the best baseline are statistically significant ($p\text{-value}< 0.05$) under one-sample t-tests.}
\label{tab:overall_per}
\begin{center}
\renewcommand\arraystretch{1.1}
\setlength{\tabcolsep}{0.6mm}{
\resizebox{\textwidth}{!}{
\begin{tabular}{l|cccc|cccc|cccc}
\hline
 & \multicolumn{4}{c|}{\textbf{Yelp}} & \multicolumn{4}{c|}{\textbf{Book}} & \multicolumn{4}{c}{\textbf{Electronics}} \\ 
Methods & R@10 & R@20 & N@10 & N@20 & R@10 & R@20 & N@10 & N@20 & R@10 & R@20 & N@10 & N@20 \\ \hline
MF & 0.0385 & 0.0659 & 0.0269 & 0.0365 & 0.0206 & 0.0355 & 0.0142 & 0.0193 & 0.0333 & 0.0551 & 0.0186 & 0.0243 \\
LightGCN & 0.0402 & 0.0695 & 0.0288 & 0.0390 & 0.0252 & 0.0434 & 0.0170 & 0.0233 & 0.0365 & 0.0560 & 0.0194 & 0.0246 \\ \hline
MultiVAE & 0.0427 & 0.0728 & 0.0303 & 0.0409 & 0.0280 & 0.0475 & 0.0197 & 0.0264 & 0.0419 & 0.0658 & 0.0224 & 0.0286 \\
MacridVAE & {\ul 0.0442} & {\ul 0.0770} & 0.0319 & {\ul 0.0434} & 0.0409 & 0.0667 & 0.0288 & 0.0378 & 0.0424 & 0.0654 & 0.0246 & 0.0306 \\ \hline
{DIB} & {0.0375} & {0.0654} & {0.0264} & {0.0362} & {0.0211} & {0.0362} & {0.0144} & {0.0196} & {0.0319} & {0.0530} & {0.0176} & {0.0231} \\
{COR} & {0.0411} & {0.0690} & {0.0293} & {0.0392} & {0.0400} & {0.0681} & {0.0286} & {0.0385} & {0.0428} & {0.0625} & {0.0237} & {0.0289} \\ \hline
{DIEN} & {0.0275} & {0.0449} & {0.0202} & {0.0263} & {0.0279} & {0.0382} & {0.0250} & {0.0284} & {0.0347} & {0.0512} & {0.0196} & {0.0240} \\
{MGS} & {0.0423} & {0.0696} & {0.0313} & {0.0409} & {0.0515} & {0.0738} & {0.0458} & {0.0532} & {0.0415} & {0.0625} & {0.0230} & {0.0285}\\
DSSRec & 0.0413 & 0.0697 & 0.0299 & 0.0400 & 0.0539 & 0.0790 & 0.0448 & 0.0534 & 0.0503 & {\ul 0.0780} & 0.0269 & 0.0343 \\
ACVAE & 0.0439 & 0.0750 & {\ul 0.0322} & 0.0432 & {\ul 0.0563} & {\ul 0.0860} & {\ul 0.0477} & {\ul 0.0576} & {\ul 0.0510} & 0.0766 & {\ul 0.0290} & {\ul 0.0359} \\
CauseRec & 0.0433 & 0.0762 & 0.0300 & 0.0417 & 0.0484 & 0.0753 & 0.0391 & 0.0482 & 0.0445 & 0.0744 & 0.0230 & 0.0309 \\ \hline 
CDR & \textbf{0.0528*} & \textbf{0.0880*} & \textbf{0.0392*} & \textbf{0.0518*} & \textbf{0.0721*} & \textbf{0.1042*} & \textbf{0.0598*} & \textbf{0.0708*} & \textbf{0.0647*} & \textbf{0.0933*} & \textbf{0.0373*} & \textbf{0.0449*} \\
\% Improve. & 19.46\% & 14.29\% & 21.74\% & 19.35\% & 28.06\% & 21.16\% & 25.37\% & 22.92\% & 26.86\% & 19.62\% & 28.62\% & 25.07\% \\\hline
\end{tabular}
}
}
\end{center}
\vspace{-0.1cm}
\end{table*}

\subsection{Overall Performance (RQ1)}\label{sec:overall_per}
We present the results of the baselines and CDR on the three datasets in Table \ref{tab:overall_per}. From the table, we have the following observations:

\begin{itemize}[leftmargin=*]
    \item MacridVAE consistently outperforms MF, LightGCN, and MultiVAE on the three datasets. We attribute the superior performance to the disentangled user representations of MacridVAE. The preference shifts only affect partial user representations while most disentangled user representations of MacridVAE are robust to the shifts. Besides, the sequential models (\ie DSSRec, ACVAE, and CauseRec) {and the session-based model (\ie MGS)} usually perform better than MF, LightGCN, and MultiVAE, which verifies the effectiveness of considering temporal information in capturing preference shifts.
    
    \item {The performance of MacridVAE is better than that of sequential models (DSSRec, ACVAE, and CauseRec) on Yelp while the sequential models surpass MacridVAE on Book and Electronics. Meanwhile, DIEN outperforms MF on Book and Electronics while yields inferior performance on Yelp.}
    This is because the effect of preference shifts is quite different on the three datasets. 
    Temporal preference shifts are stronger on Book and Electronics, and thus sequential modeling is more effective to capture the shifts. In contrast, user preference over food is relatively stable on Yelp, where MacridVAE is superior to model the invariant preference. 
    
    \item In sequential models, ACVAE usually achieves higher performance than DSSRec and CauseRec. This is probably because ACVAE introduces adversarial training and contrastive learning into sequential VAE, which also encourages the independence of latent factors in user representations. Such independence might have a similar effect as disentangled representations of CDR, \ie the sparse structure. The main difference is that the disentangled representations of CDR are learned from multiple environments, which are more robust under preference shifts. Besides, the inferior performance of DSSRec and CauseRec might be attributed to the improper intention clustering~\cite{ma2020disentangled} and inaccurate identification of dispensable concepts~\cite{zhang2021cause}, respectively. 
    
   \item {COR is usually comparable with MacridVAE on Book and Electronics while performs worse than sequential models such as ACVAE. This is reasonable since COR eliminates the out-of-date information and reuses the stable preference, leading to robust user representations against user preference shifts. However, COR ignores the temporal feature shifts, resulting in worse performance than sequential models. Besides, DIB shows relatively worse results, which is possible because that biased embeddings might be still useful and totally discarding them loses critical user preference.}  
    
    \item CDR significantly yields the best performance on the three datasets. Specifically, the performance improvements of CDR over the best baseline \wrt Recall@10 are 19.46\%, 28.06\%, and 26.86\% on Yelp, Book, and Electronics, respectively. This justifies the superiority of handling user preference shifts via the CDR framework. CDR does not only capture the temporal preference trend between environments for better preference estimation, but also learns a robust structure from user preference to interactions, leading to better interaction prediction. 

\end{itemize}

\begin{figure}[t]
\setlength{\abovecaptionskip}{0.1cm}
\setlength{\belowcaptionskip}{0cm}
  \centering 
  \hspace{-0.1in}
  \subfigure{
    \includegraphics[width=1.8in]{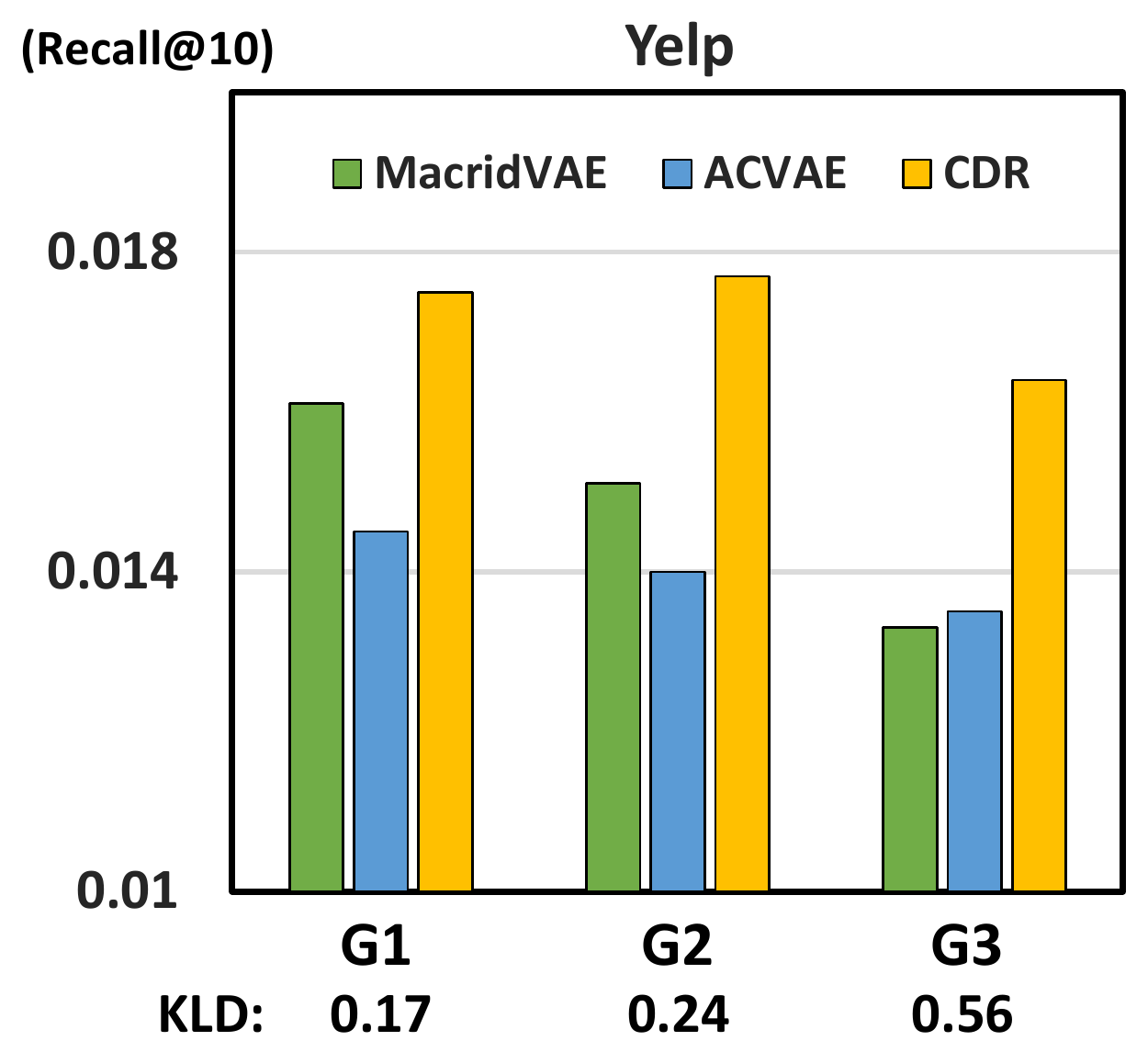}} 
   \hspace{-0.1in}
  \subfigure{
    \includegraphics[width=1.8in]{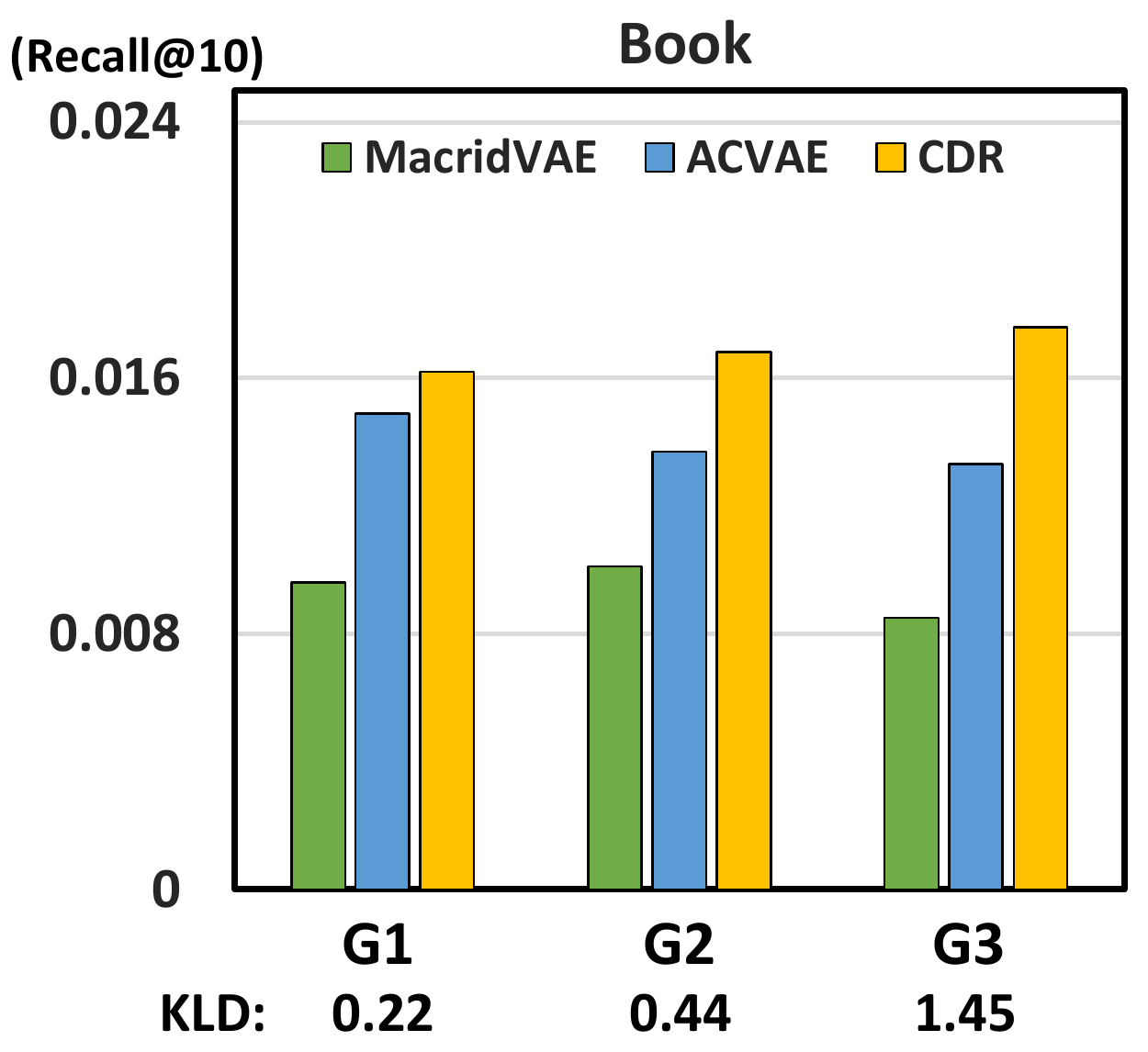}} 
   \hspace{-0.1in}
  \subfigure{
    \includegraphics[width=1.8in]{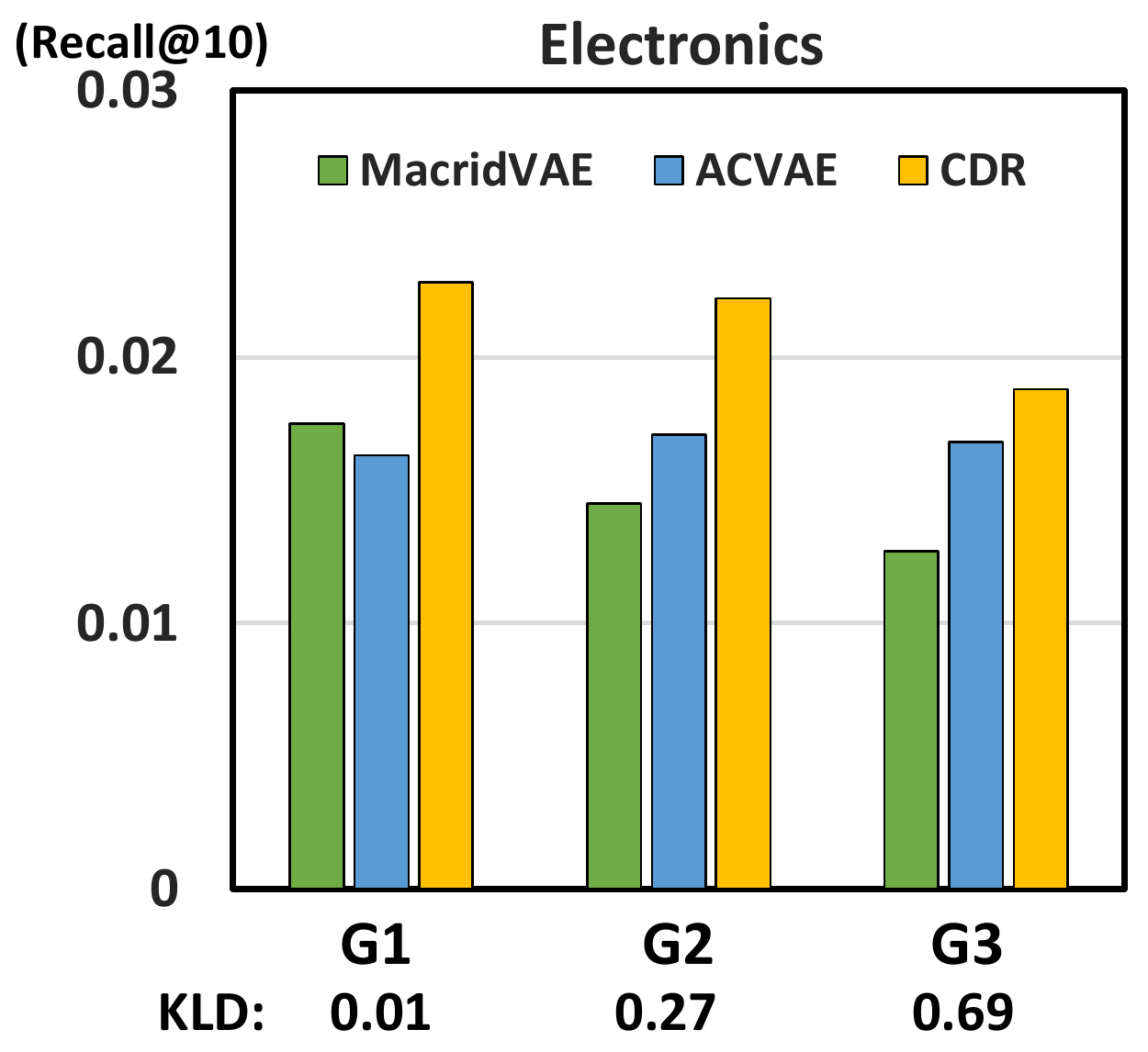}} 
  \hspace{-0.1in}
  \caption{{Performance comparison on three user groups with different strengths of shifts, where the shifts increase from G1, G2, to G3.}}
  \label{fig:user_cmp}
\end{figure}

{To evaluate the performance of CDR under different strengths of shifts, we split users into groups according to the KL divergence between training and testing environments \wrt their interacted item categories. As shown in Figure~\ref{fig:user_cmp}, the preference shifts increase from G1, G2, to G3. From the figure, we can find that 1) CDR consistently achieves better performance across three groups; and 2) the performance of the best baselines, MacridVAE and ACVAE, usually decreases in the G3 group with large shifts while CDR still shows large improvements over the baselines, validating its stronger OOD generalization ability.}

\begin{figure*}[t]
\setlength{\abovecaptionskip}{0cm}
\setlength{\belowcaptionskip}{0cm}
  \centering 
  \subfigure{
    \includegraphics[width=1.35in]{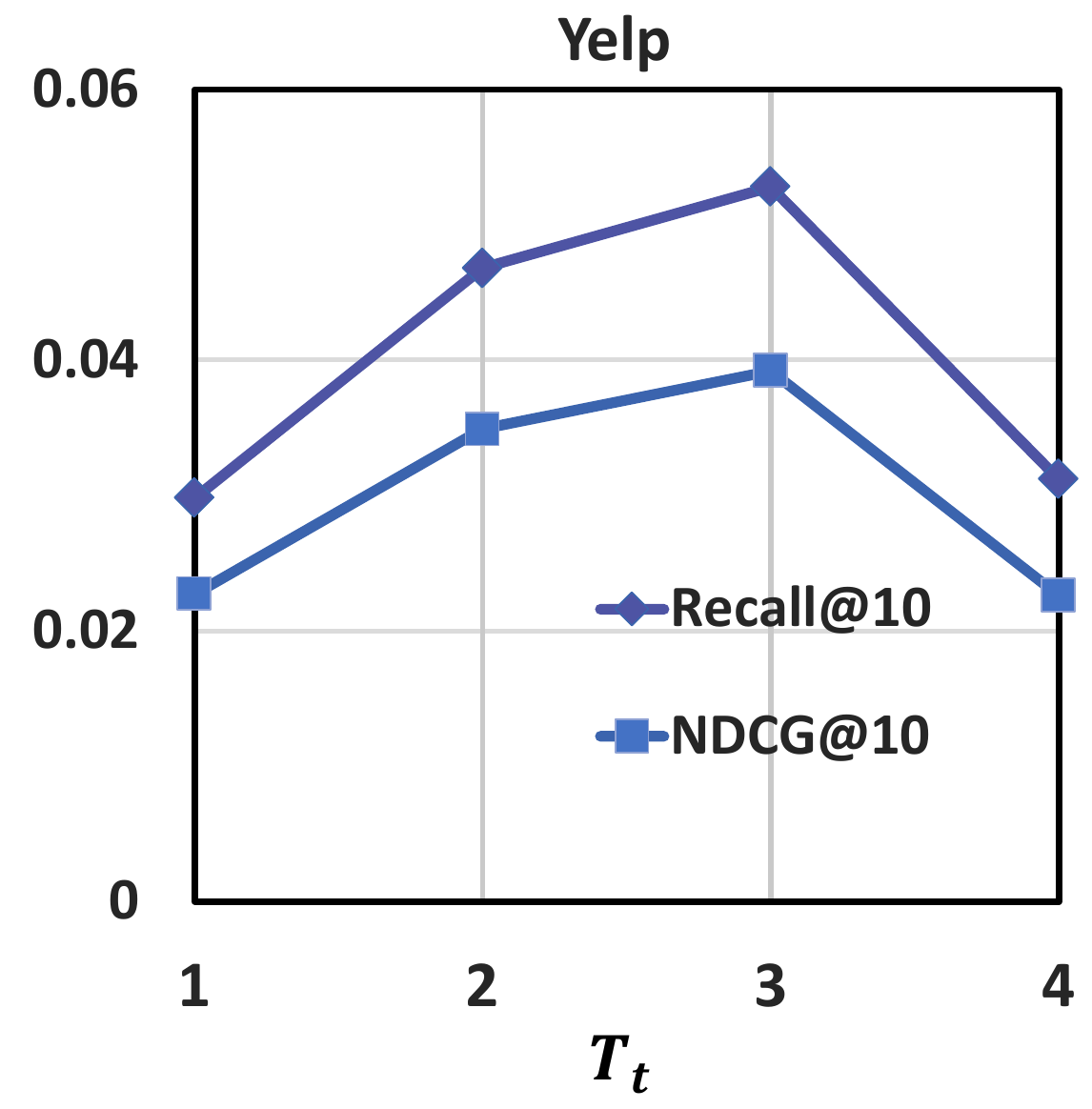}} 
  \subfigure{
    \includegraphics[width=1.35in]{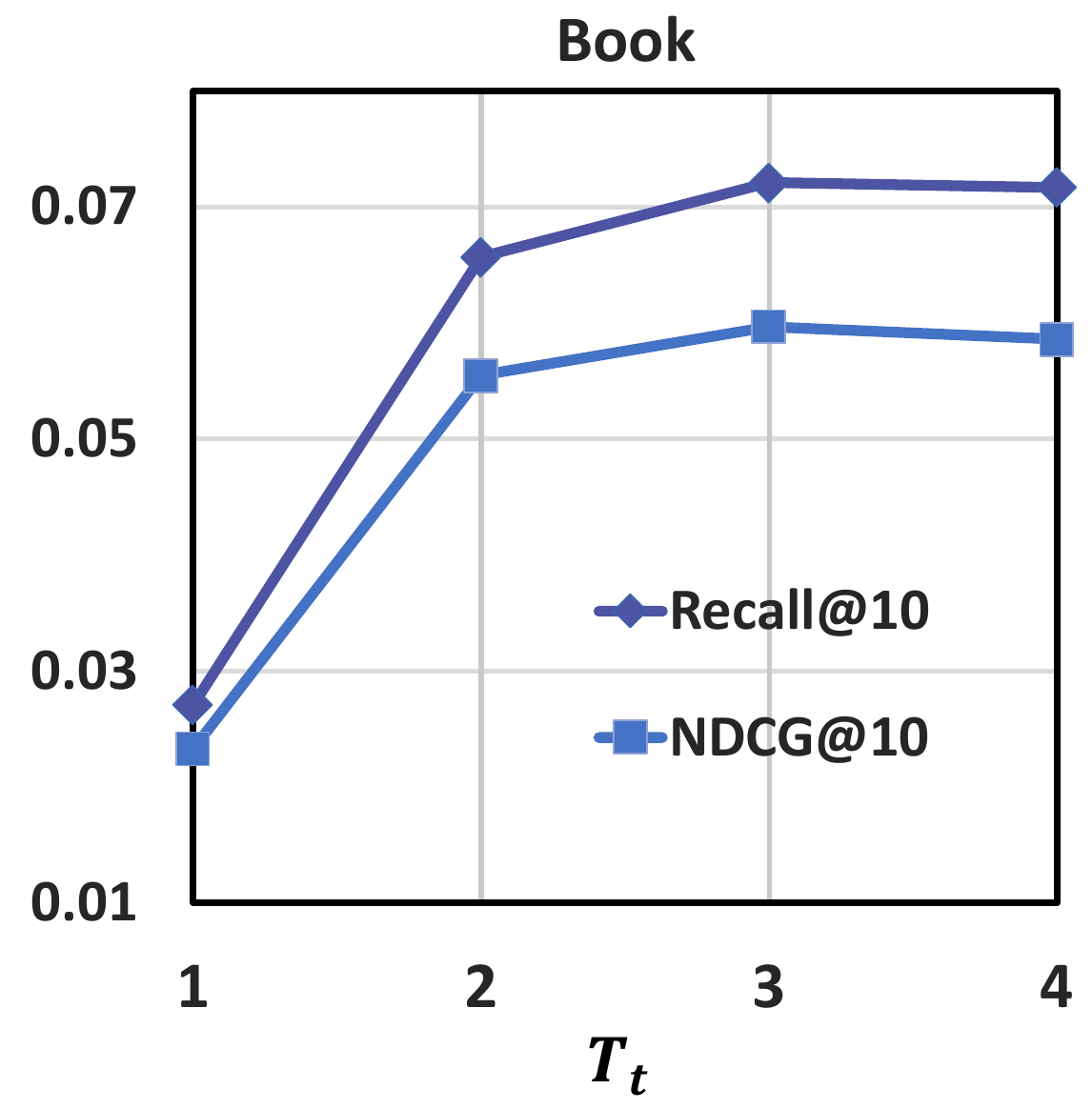}} 
  \subfigure{
    \includegraphics[width=1.35in]{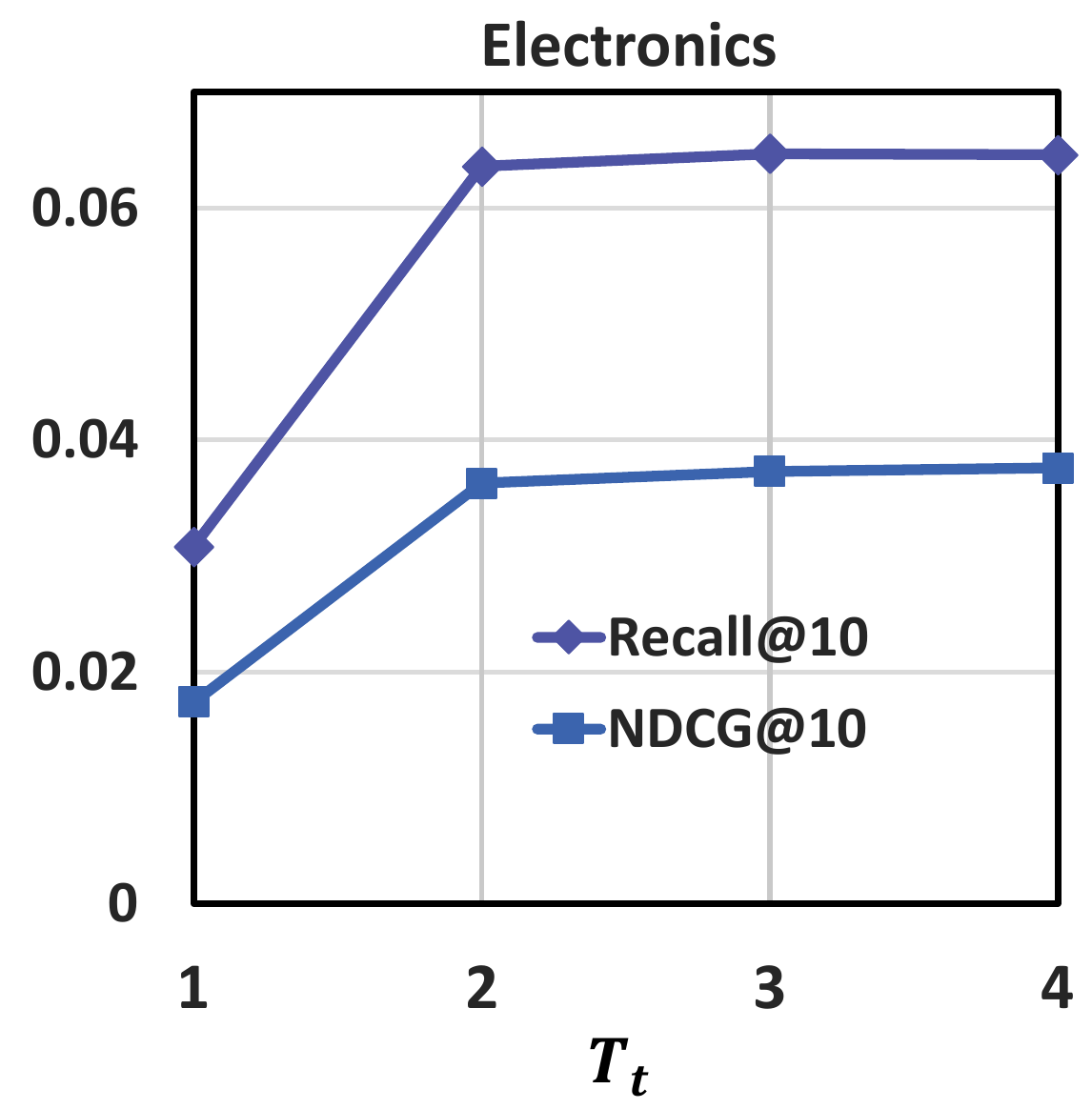}} 
  \subfigure{
    \includegraphics[width=1.35in]{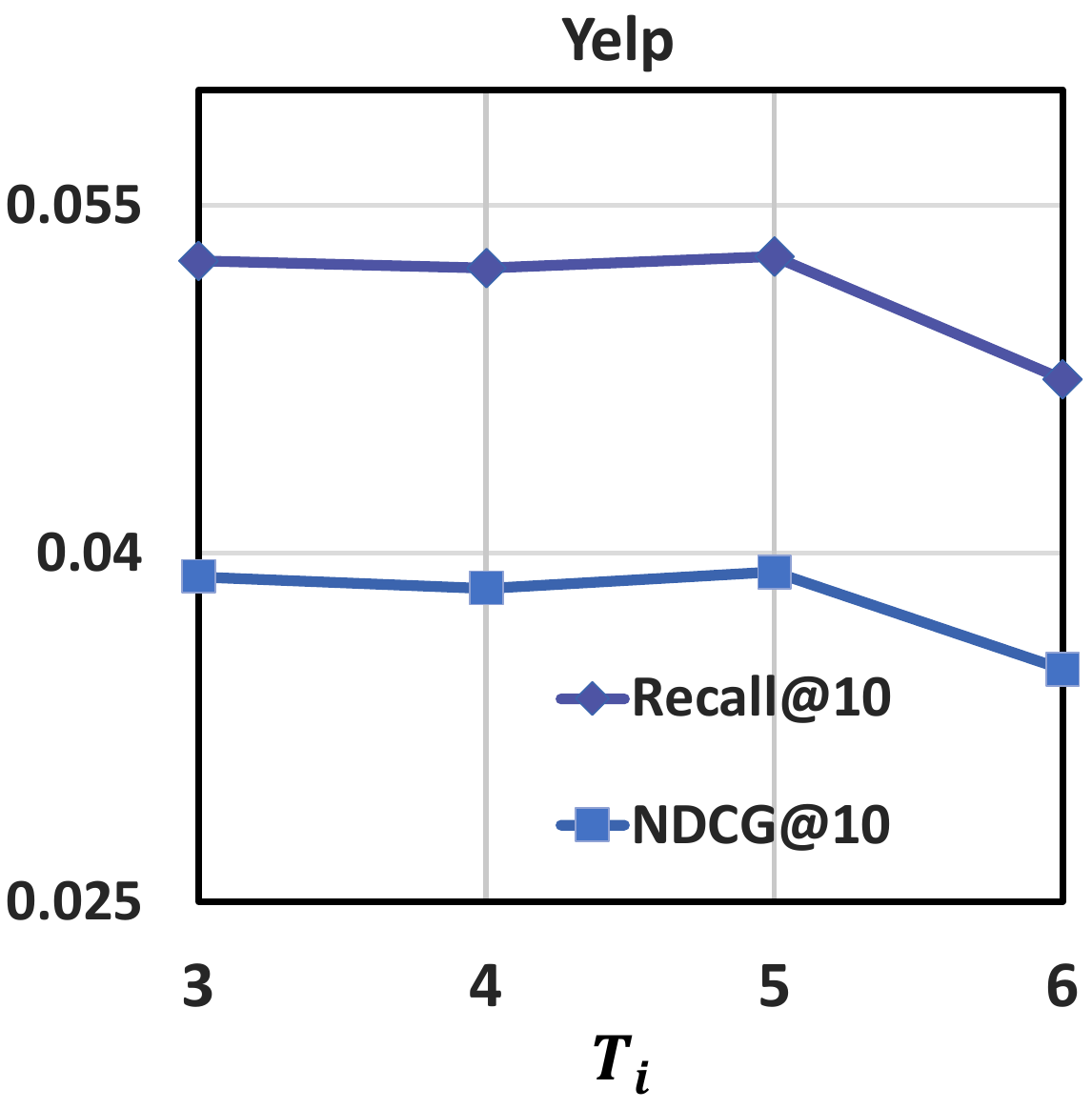}} 
  \subfigure{
    \includegraphics[width=1.35in]{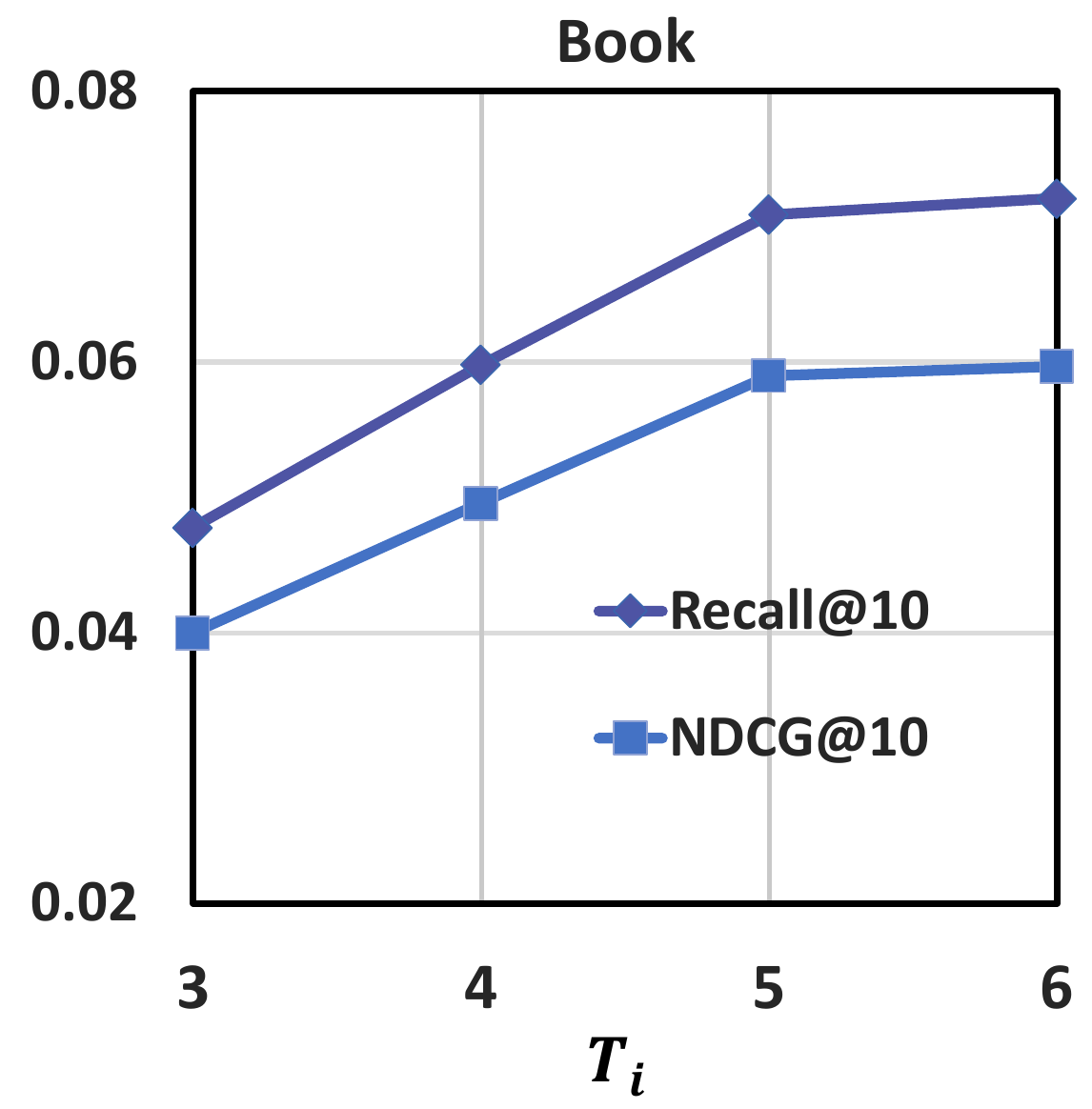}} 
  \subfigure{
    \includegraphics[width=1.35in]{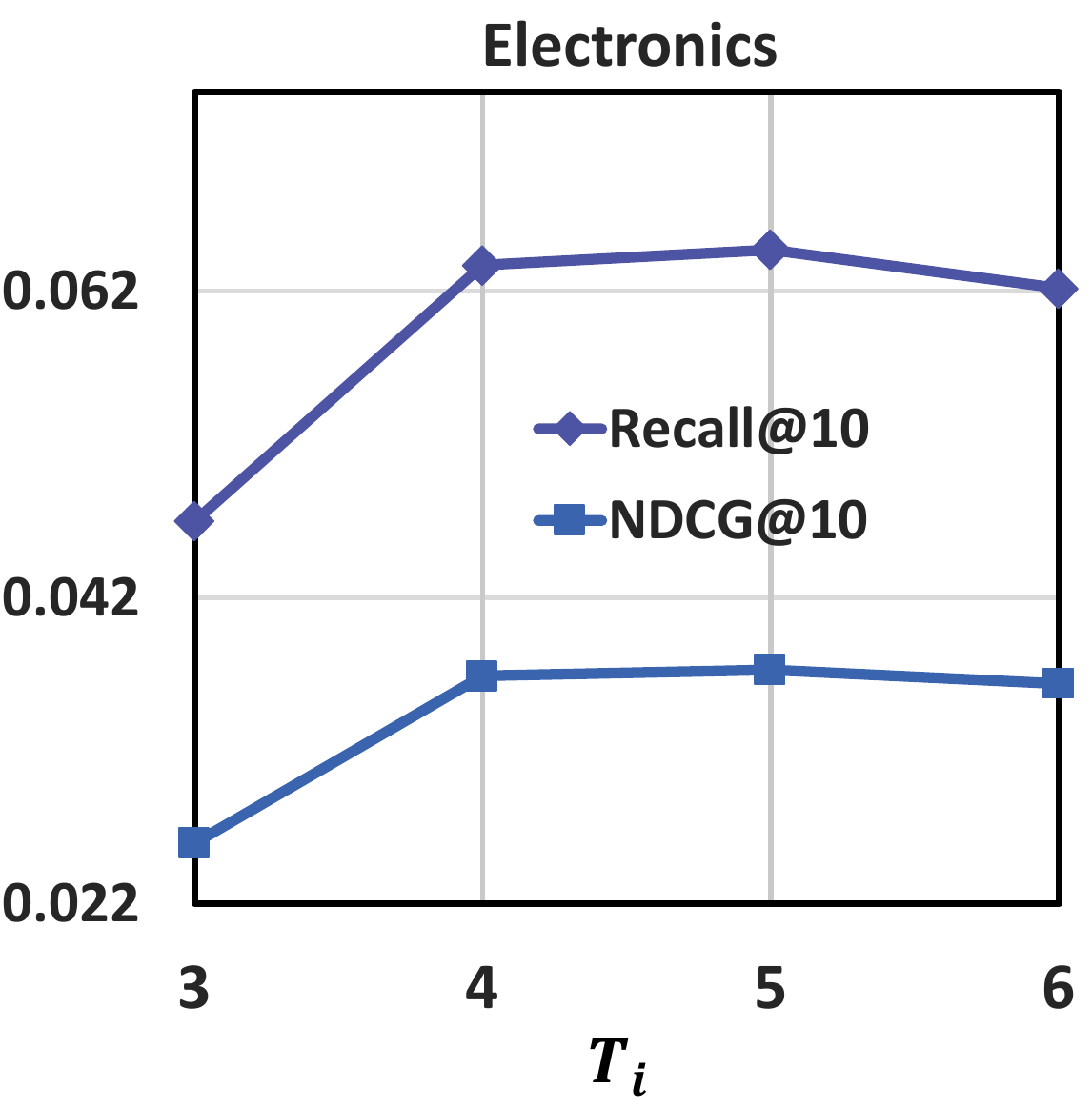}} 
  \caption{{Effect of environment numbers in the training ($T_t$) and inference ($T_i$) stages. We obtain the results of different $T_i$ by using the best $T_t=3$ for training and vary $T_i$ for inference.}
  }
  \label{fig:envi_num}
  \vspace{-0.5cm}
\end{figure*}

\subsection{{In-depth Analysis (RQ2)}}
In addition to overall performance comparison, we conduct the in-depth analysis to study the effectiveness of different components in CDR, including multiple environments, the sparse structure for the disentanglement, conditional relations, inference strategies, the multi-objective loss, and hyper-parameter settings. Lastly, we provide some cases to show the effectiveness of CDR at a fine-grained level. 

\begin{figure}[t]
\setlength{\abovecaptionskip}{0.1cm}
\setlength{\belowcaptionskip}{0cm}
  \centering 
  \hspace{-0.1in}
  \subfigure{
    \includegraphics[width=1.8in]{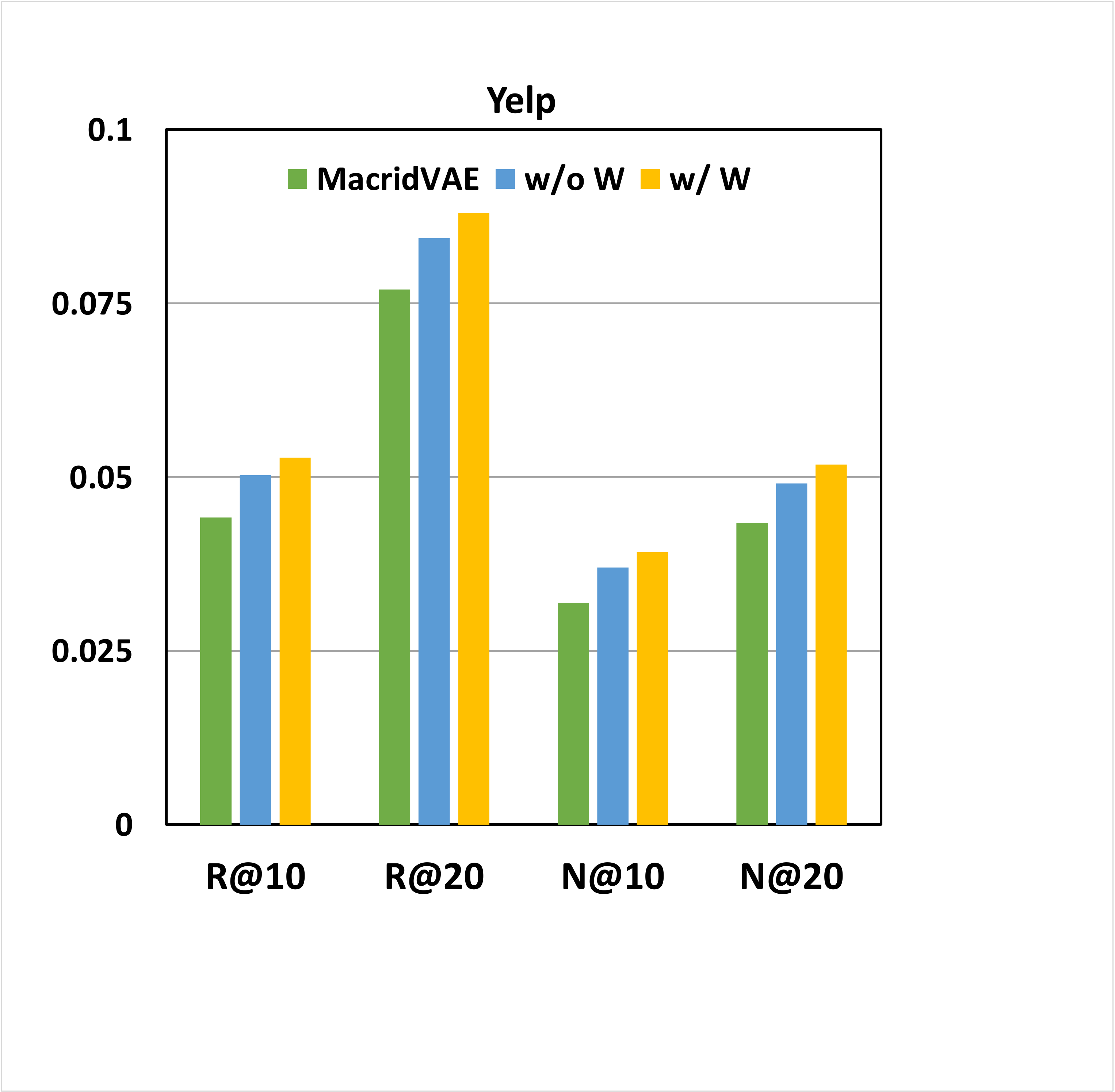}} 
   \hspace{-0.1in}
  \subfigure{
    \includegraphics[width=1.8in]{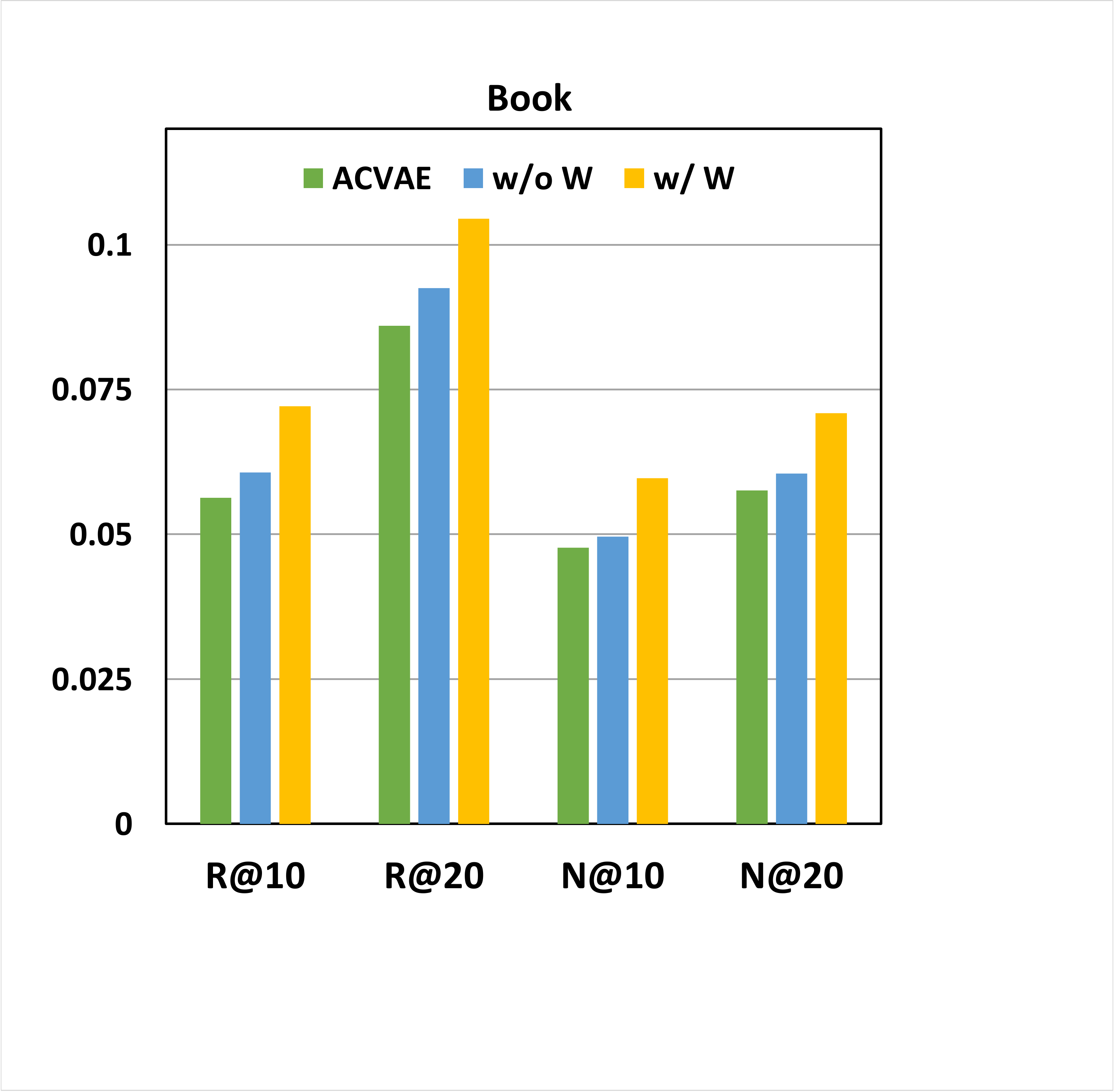}} 
   \hspace{-0.1in}
  \subfigure{
    \includegraphics[width=1.8in]{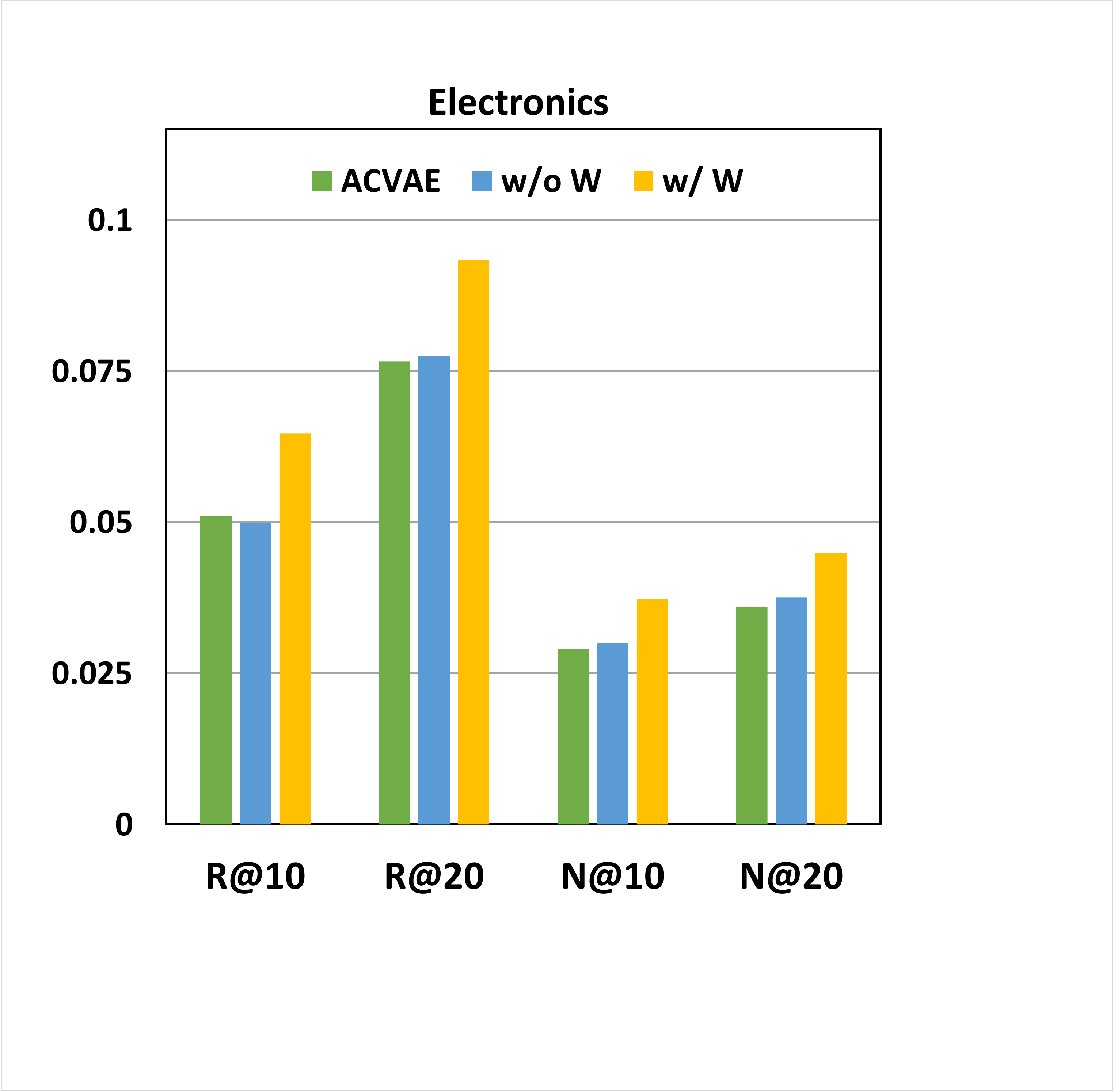}} 
  \hspace{-0.1in} 
  \caption{Ablation study of the sparse structure (\ie $\bm{W}_z$ and $\bm{W}_x$). We only show the best baseline on each dataset to save space.}
  \label{fig:ablation_W}
\end{figure}

\subsubsection{\textbf{Effect of Multiple Environments}}
As illustrated in Section \ref{sec:environment}, the choice of the environment number $T$ is essential. 
As such, we study the effect of multiple environments by varying the environment numbers in the training and inference periods. We report the results with different environment numbers during training ($T_t$) and inference ($T_i$) in Figure \ref{fig:envi_num}. From the figure, we have the following findings.
\begin{itemize}[leftmargin=*]
    \item During training, the performance rises at first, and then drops with the increase of $T_t$. The rise validates the effectiveness of learning user representations from multiple environments instead of one environment. Besides, the performance drop also verifies the arguments in Section \ref{sec:environment}: more environments will make the interactions in each environment sparse, hindering the learning of invariant user preference within an environment and the disentangled representations (\ie the sparse structure $\bm{W}_z$ and $\bm{W}_x$). 
    
    \item In the inference stage, a larger $T_i$ than $T_t$ is able to further improve the performance of CDR, especially on Book and Electronics. This is due to the better utilization of temporal information. During inference, the sparsity issue in each environment will not affect the optimization of CDR. As such, we can fully utilize the temporal information of interactions and consider more fine-grained temporal preference shifts by using larger $T_i$ for the inference. 
    
    \item The effectiveness of increasing $T_i$ is more significant on Book and Electronics than that of Yelp. The underlying reason is that Book and Electronics have stronger temporal shifts as discussed in Section \ref{sec:overall_per} and more environments help to capture more fine-grained shifts between environments. In contrast, user preference is relatively stable on Yelp, and thus CDR leverages fewer environments to better capture the invariant preference within each environment. This shows that CDR can flexibly balance the learning of invariant and shifted preference on different datasets by adjusting the environment number. 
\end{itemize}


\subsubsection{\textbf{Effect of Sparse Structure}}
To validate the effectiveness of the sparse structure from user preference to interactions, we perform the ablation studies over the two matrices $\bm{W}_z$ and $\bm{W}_x$. 
The results with (\ie w/ $\bm{W}$) and without (\ie w/o $\bm{W}$) the two matrices are provided in Figure \ref{fig:ablation_W}. The ablation of $\bm{W}_z$ and $\bm{W}_x$ denotes that CDR only uses an MLP model $f_\gamma(\cdot)$ to obtain $f_{\theta_2}(\bm{z}_t)$ in Eq. (\ref{eqn:f_theta_2}). From Figure \ref{fig:ablation_W}, we can observe that: 
\begin{itemize}[leftmargin=*]
    \item The performance declines if the two matrices are removed, showing the effectiveness of the sparse structure in modeling the effect of user preference shifts.
    \item CDR without $\bm{W}_z$ and $\bm{W}_x$ still outperforms the best baselines, \ie MacridVAE and ACVAE. Such improvements are attributed to the division of environments: without the disentanglement via the sparse structure, CDR still captures both the preference shifts between environments and the invariant preference within each environment by following the robust causal relations~\cite{locatello2020weakly, yoshua2020a}. 
\end{itemize}


\begin{table}[t]
\setlength{\abovecaptionskip}{0cm}
\setlength{\belowcaptionskip}{0cm}
\caption{{Performance comparison of CDR with and without conditional relations.}}
\label{tab:conditional_cmp}
\begin{center}
\setlength{\tabcolsep}{2.8mm}{
\resizebox{0.7\textwidth}{!}{
\begin{tabular}{lcccccc}
\toprule
 & \multicolumn{2}{c}{Yelp} & \multicolumn{2}{c}{Book} & \multicolumn{2}{c}{Electronics} \\
 & R@10 & N@10 & R@10 & N@10 & R@10 & N@10 \\ \hline
ACVAE & 0.0439 & 0.0322 & 0.0563 & 0.0477 & 0.0510 & 0.0290 \\
CDR with $E_{t-1}\rightarrow E_t$ & 0.0450 & 0.0335 & 0.0584 & 0.0466 & 0.0549 & 0.0306 \\
CDR with $X_{t-1}\rightarrow Z_{t}$ & 0.0458 & 0.0340 & 0.0572 & 0.0491 & 0.0553 & 0.0314\\
Vanilla CDR & 0.0528 & 0.0392 & 0.0721 & 0.0598 & 0.0647 & 0.0373 \\\bottomrule
\end{tabular}
}
}
\end{center}
\vspace{-0.2cm}
\end{table}

\subsubsection{{\textbf{Effect of Conditional Relations}}}\label{sec:condition_rel}
{
We conduct experiments to compare the CDR performance with and without considering the conditional relations of $E_{t-1}\rightarrow E_{t}$ and $X_{t-1}\rightarrow Z_{t}$. Considering them will change the encoder $q(\bm{e}_t|\bm{x}_t)$ and the decoder module $p(\bm{z}_t|\bm{e}_t, \bm{z}_{t-1})$ into $q(\bm{e}_t|\bm{x}_t, \bm{e}_{t-1})$ and $p(\bm{z}_t|\bm{e}_t, \bm{z}_{t-1}, \bm{x}_{t-1})$, respectively. 
Such changes introduce more parameters due to the larger input dimension. 
The experimental results are presented in Table~\ref{tab:conditional_cmp}, from which we observe that 1) CDR with $E_{t-1}\rightarrow E_{t}$ or $X_{t-1}\rightarrow Z_{t}$ has inferior performance than the vanilla CDR. The possible reasons are that these conditional relations are not strong over a large proportion of users, and meanwhile the CDR with more parameters might overfit the training data, hurting the generalization ability in a new environment. Besides, 2) CDR with $E_{t-1}\rightarrow E_{t}$ or $X_{t-1}\rightarrow Z_{t}$ still surpasses the best baseline ACVAE, validating the effectiveness of modeling preference shifts and sparse influence by this CDR framework. 
}

\begin{table}[t]
\setlength{\abovecaptionskip}{0cm}
\setlength{\belowcaptionskip}{0cm}
\caption{{Performance comparison of three inference strategies.}}
\label{tab:infer_cmp}
\begin{center}
\setlength{\tabcolsep}{2.8mm}{
\resizebox{0.66\textwidth}{!}{
\begin{tabular}{lcccccc}
\toprule
 & \multicolumn{2}{c}{Yelp} & \multicolumn{2}{c}{Book} & \multicolumn{2}{c}{Electronics} \\
 & R@10 & N@10 & R@10 & N@10 & R@10 & N@10 \\ \hline
1)  $\bm{z}_T$ & 0.0528 & 0.0392 & 0.0721 & 0.0597 & 0.0647 & 0.0373 \\
2) \text{avg} $\bm{x}_{1:T}$ & 0.0429 & 0.0312 & 0.0288 & 0.0206 & 0.0476 & 0.0258 \\
3) $\bm{z}_{T+1}$ & 0.0470 & 0.0340 & 0.0550 & 0.0449 & 0.0613 & 0.0336 \\\bottomrule
\end{tabular}
}
}
\end{center}
\vspace{-0.2cm}
\end{table}

\subsubsection{{\textbf{Effect of Inference Strategies}}}\label{sec:infer_unk}

{In Table~\ref{tab:infer_cmp}, we report the results of the three inference strategies detailed in Section~\ref{sec:task}. From Table~\ref{tab:infer_cmp}, we can find that 1) the first strategy outperforms the second and third strategies, and 2) the second one shows the worst results. These findings are reasonable because 1) the second strategy ignores the temporal distribution shifts and uniformly averages the predictions, and 2) the third strategy is better since it partly considers the shifts by $\bm{z}_T$ while $\bm{e}_{T+1}$ is still obtained by average, inevitably losing some temporal patterns. In future work, it is promising to explore more strategies to better capture temporal patterns for the inference. 
}

\begin{figure}[t]
\setlength{\abovecaptionskip}{0.10cm}
\setlength{\belowcaptionskip}{0cm}
  \centering 
  \hspace{-0.30in}
  \subfigure{
    \includegraphics[width=1.65in]{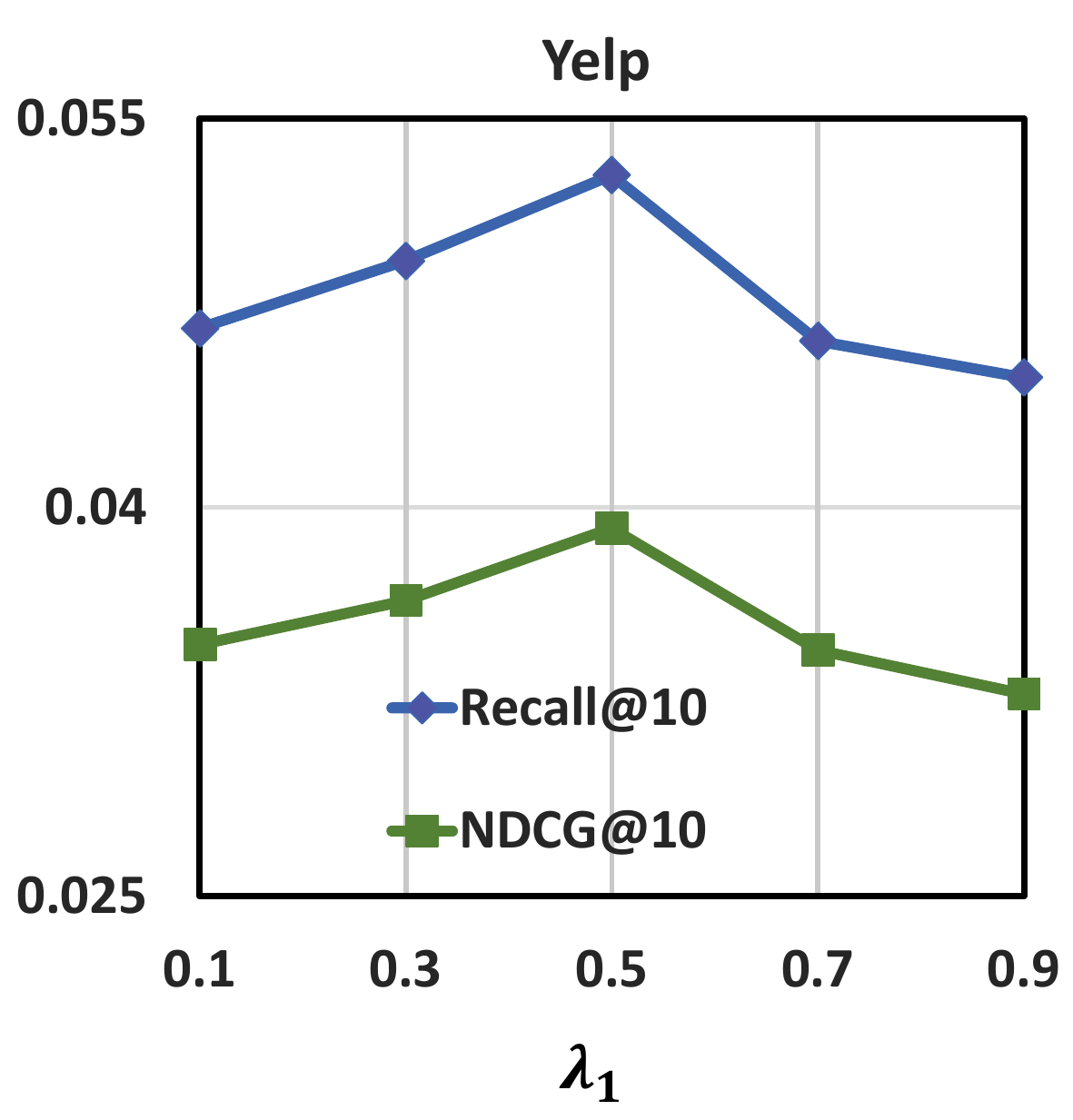}}
  \hspace{-0.1in} 
  \subfigure{
    \includegraphics[width=2in]{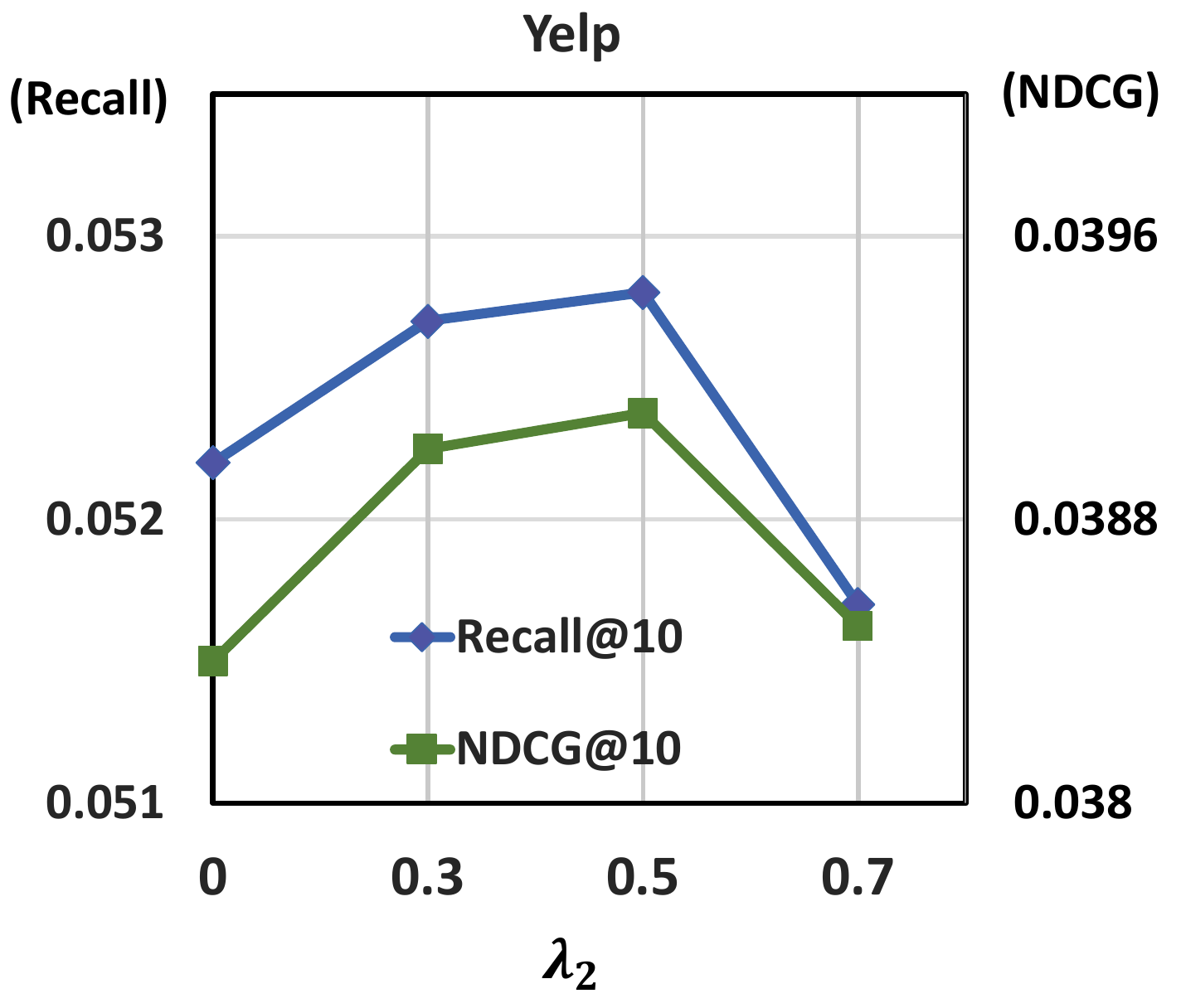}}
  \hspace{-0.1in}
  \subfigure{
    \includegraphics[width=2in]{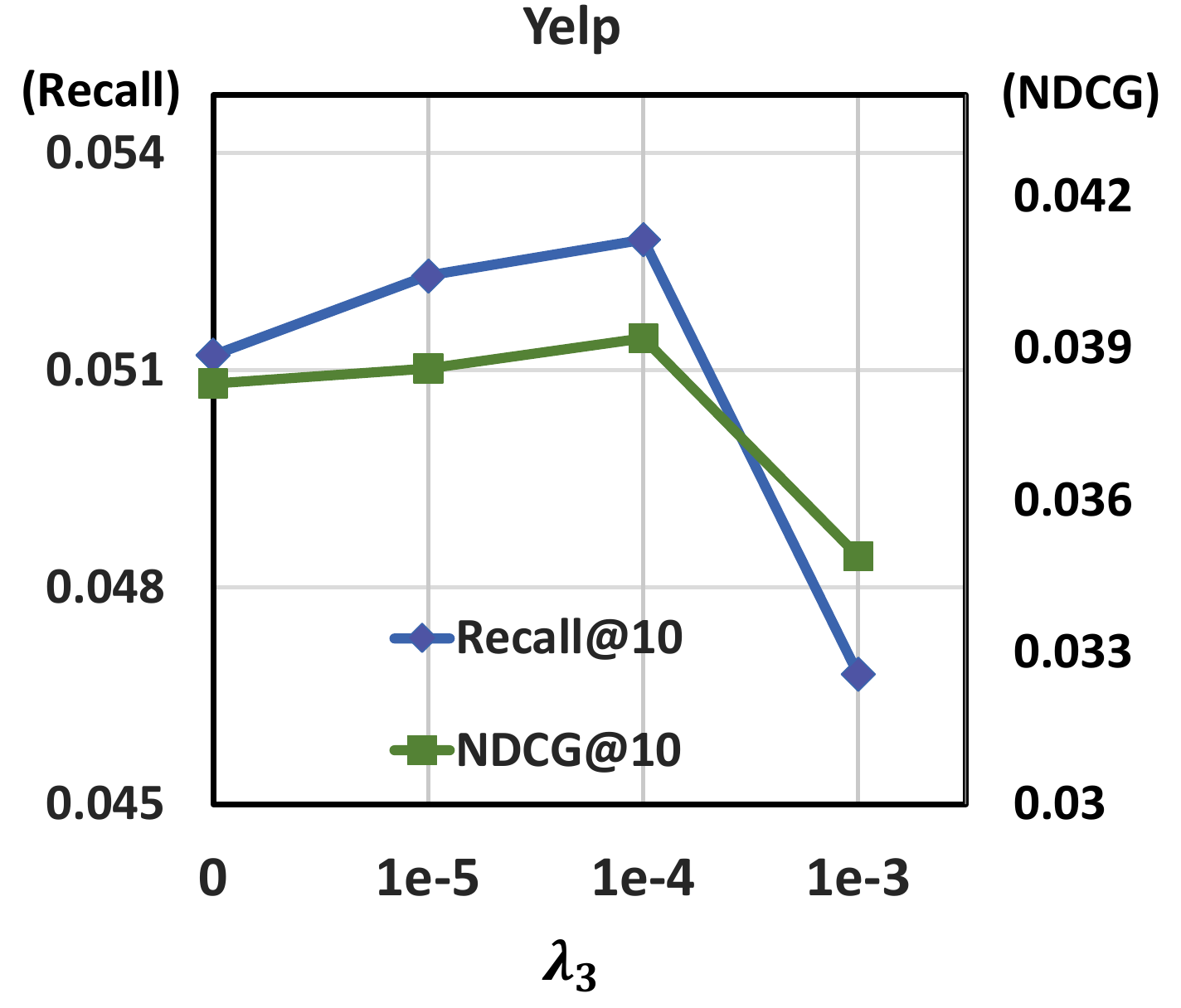}
    \hspace{-0.30in}
    }
  \subfigure{
    \hspace{-0.30in}
    \includegraphics[width=1.65in]{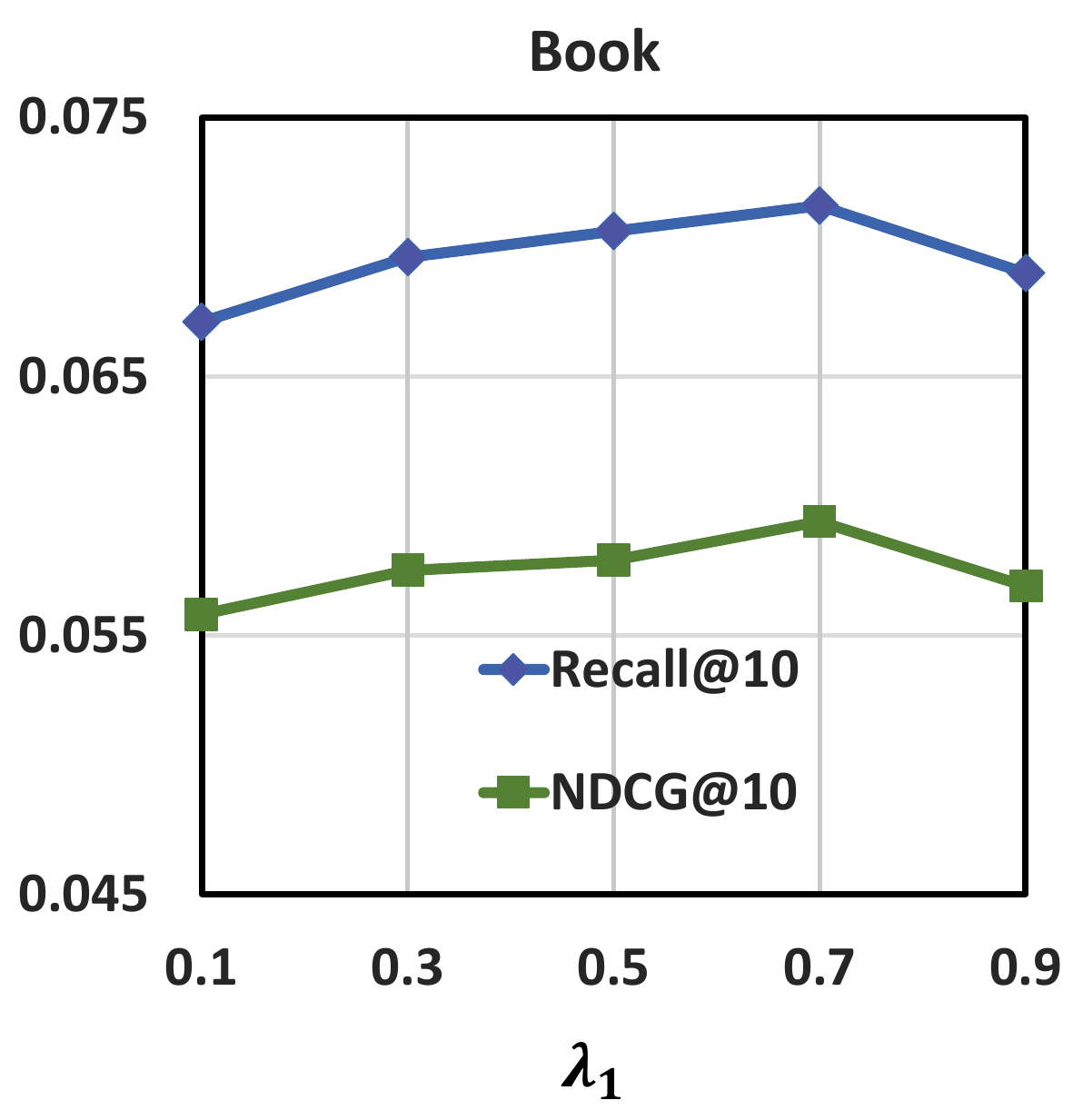}}
  \hspace{-0.10in}
  \subfigure{
    \includegraphics[width=2in]{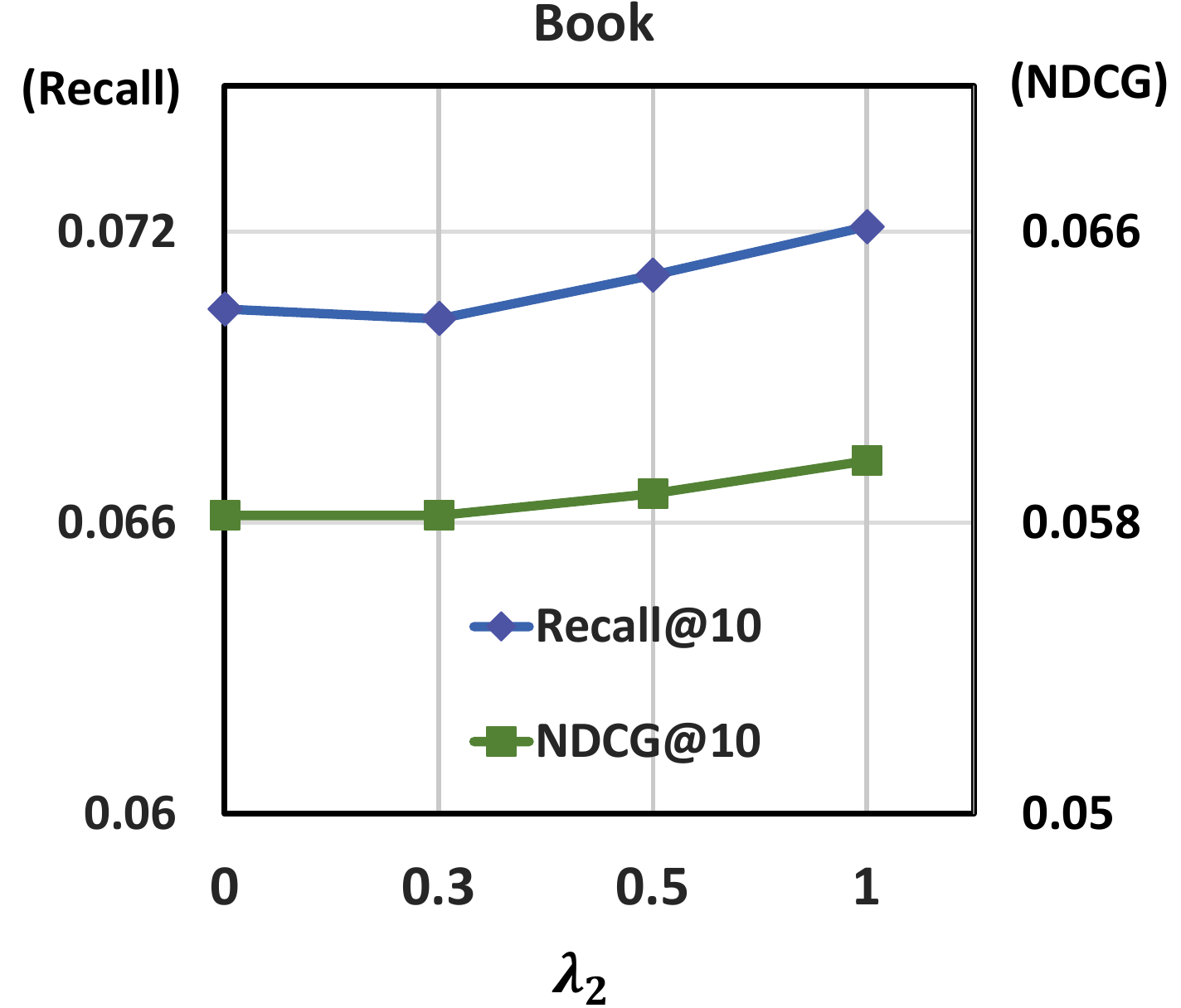}}
  \hspace{-0.10in}
  \subfigure{
    \includegraphics[width=2in]{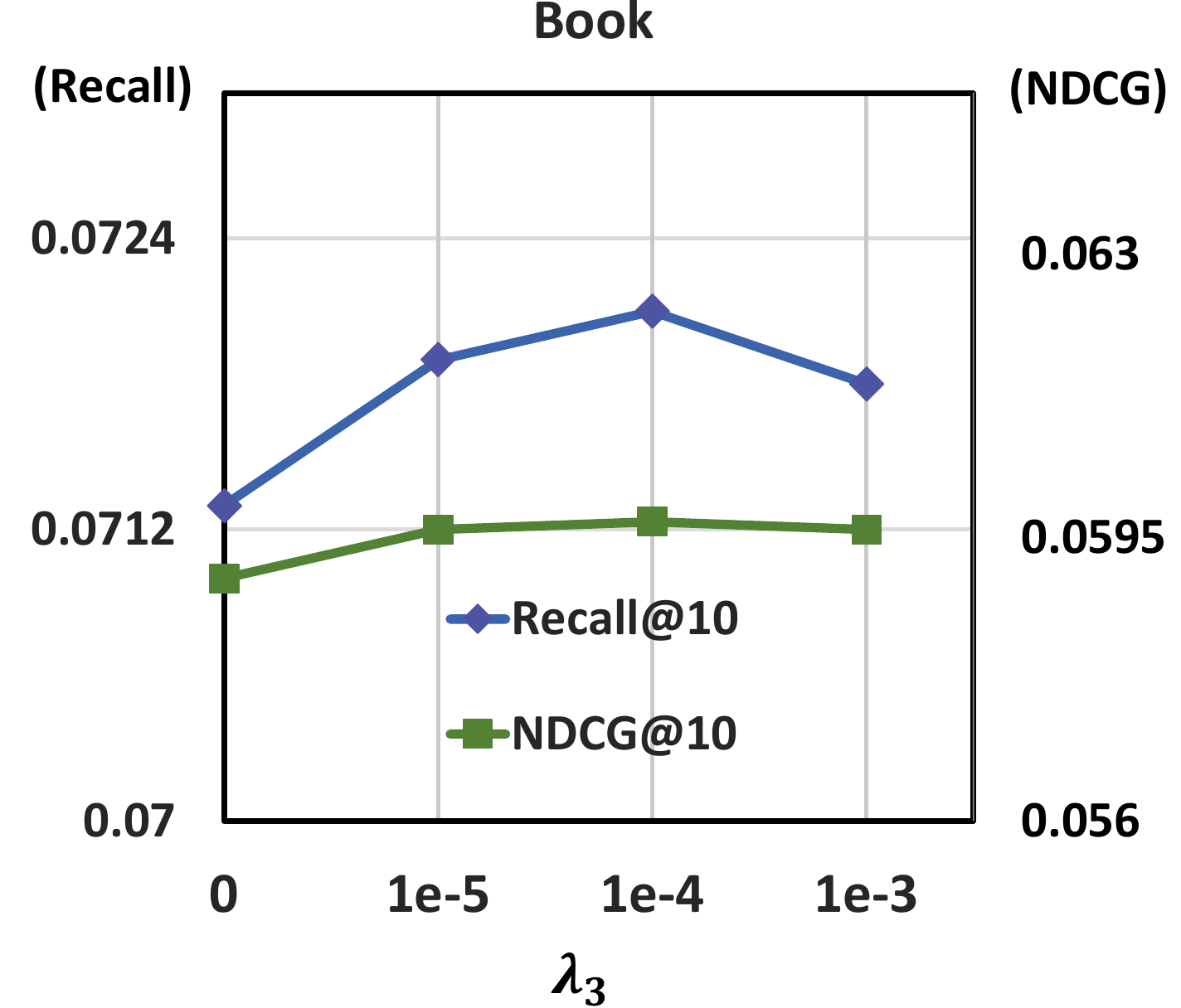}
    \hspace{-0.30in}
    }
  \caption{
    Effect of the coefficients in the multi-objective loss (\ie $\lambda_1$, $\lambda_2$, and $\lambda_3$). Specifically, $\lambda_1$ is the coefficient of KL divergence; $\lambda_2$ controls the sparsity regularization; and $\lambda_3$ adjusts the variance penalty term.
  }
  \label{fig:lambda}
\end{figure}

\subsubsection{\textbf{Effect of Multi-objective Loss}}
To analyze the influence of KL divergence, the sparsity, and variance regularizers in the multi-objective loss, we conduct experiments to compare the performance by changing their coefficients $\lambda_1$, $\lambda_2$, and $\lambda_3$. 
The results \wrt the three coefficients on Yelp and Book are presented in Figure \ref{fig:lambda}. The results on Electronics and Book are similar so that we omit the one on Electronics. From Figure \ref{fig:lambda}, we have the following observations:
\begin{itemize}[leftmargin=*]
    \item Increasing $\lambda_1$ of KL divergence is useful to improve the performance. Actually, as indicated by~\cite{higgins2017beta}, a large $\lambda_1$ for KL divergence regulates the independence of latent factors to facilitate disentangled representations; while a small $\lambda_1$ helps the model to fit the user interactions better~\cite{liang2018variational}. From Figure \ref{fig:lambda}, we find that $\lambda_1\in[0.5, 0.7]$ usually leads to a good balance and achieves superior performance, which is consistent with the findings in~\cite{liang2018variational}.
    \item The sparsity regularizer is essential to improve the performance because Recall and NDCG drop significantly when $\lambda_2$ is changed from $0.5$ to $0$. Besides, $\lambda_2$ cannot be too large, which will limit the representation capability of $\bm{W}_z$ and $\bm{W}_x$.
    \item The decreased performance from $\lambda_3=1e^{-4}$ to $\lambda_3=0$ justifies the usefulness of the variance regularizer, which balances the gradient optimization across multiple environments. Moreover, it should be noted that $\lambda_3$ is expected to be tuned in a small magnitude $[0, 1e^{-3}]$ because the stronger gradient regularization will inevitably disturb the normal optimization of parameters.
\end{itemize}

\begin{figure}[t]
\setlength{\abovecaptionskip}{0.1cm}
\centering
\includegraphics[scale=0.25]{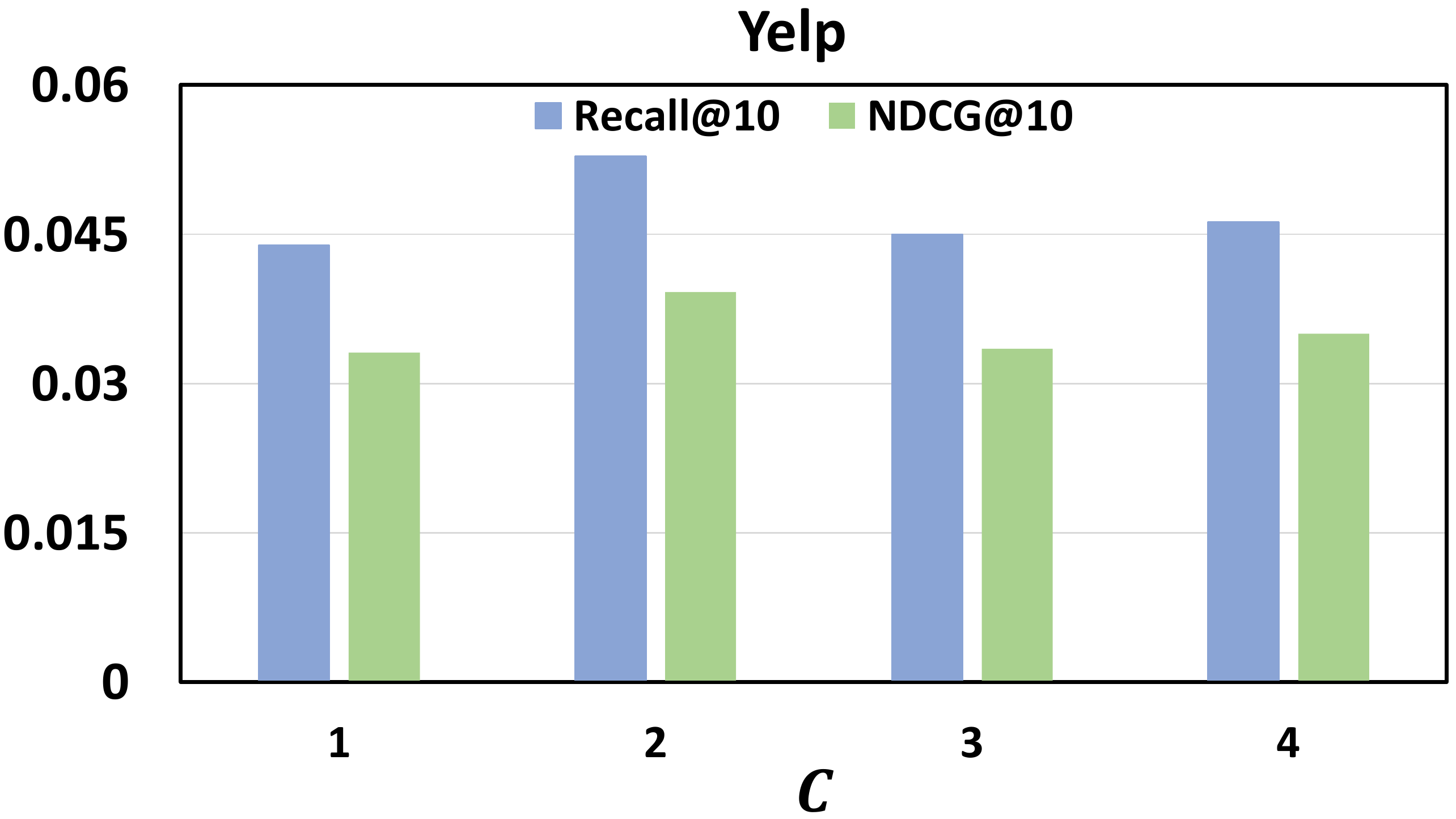}
\caption{Effect \wrt different category numbers $C$ in the structure learning.}
\label{fig:hp_C_Yelp}
\end{figure}

\subsubsection{\textbf{Effect of Category Number}}
To verify the effect of category numbers in sparse structure learning, we compare the performances with $C$ changing from 1 to 4. 
Note that $C=1$ is equivalent to removing the sparse structure, \ie using pure MLP to predict the interaction probability based on $\bm{z}_t$. 
We report the performance of CDR in Figure \ref{fig:hp_C_Yelp}.
By comparing the results \wrt Recall@10 and NDCG@10, we can find that:
\begin{itemize}[leftmargin=*]
    \item The inferior performance of $C=1$ justifies that disentangling user representations into categorical-level preference alleviates the negative effect of preference shifts. This is rational since it aligns with the real-world scenarios: items are classified into various categories and users have different preference over such categories. Once users have preference shifts, only partial factors in the user representations change and subsequently affect partial interactions, leading to better OOD generalization. 
    \item The fluctuated performance from $C=2$ to $C=4$ indicates that the performance increase is not proportional to the category number and $C=2$ shows better results. However, we usually have more categories in the real-world scenarios, for example, a variety of books. This implies that 1) it is non-trivial to recover the category-level preference from pure interaction data; and 2) incorporating item category into recommender models might help the disentanglement, which is left to future exploration.
    
\end{itemize}

\begin{figure}[t]
\setlength{\abovecaptionskip}{0cm}
\setlength{\belowcaptionskip}{-0.1cm}
  \centering 
  \hspace{-0.7in}
  \subfigure{
    \includegraphics[height=1.6in]{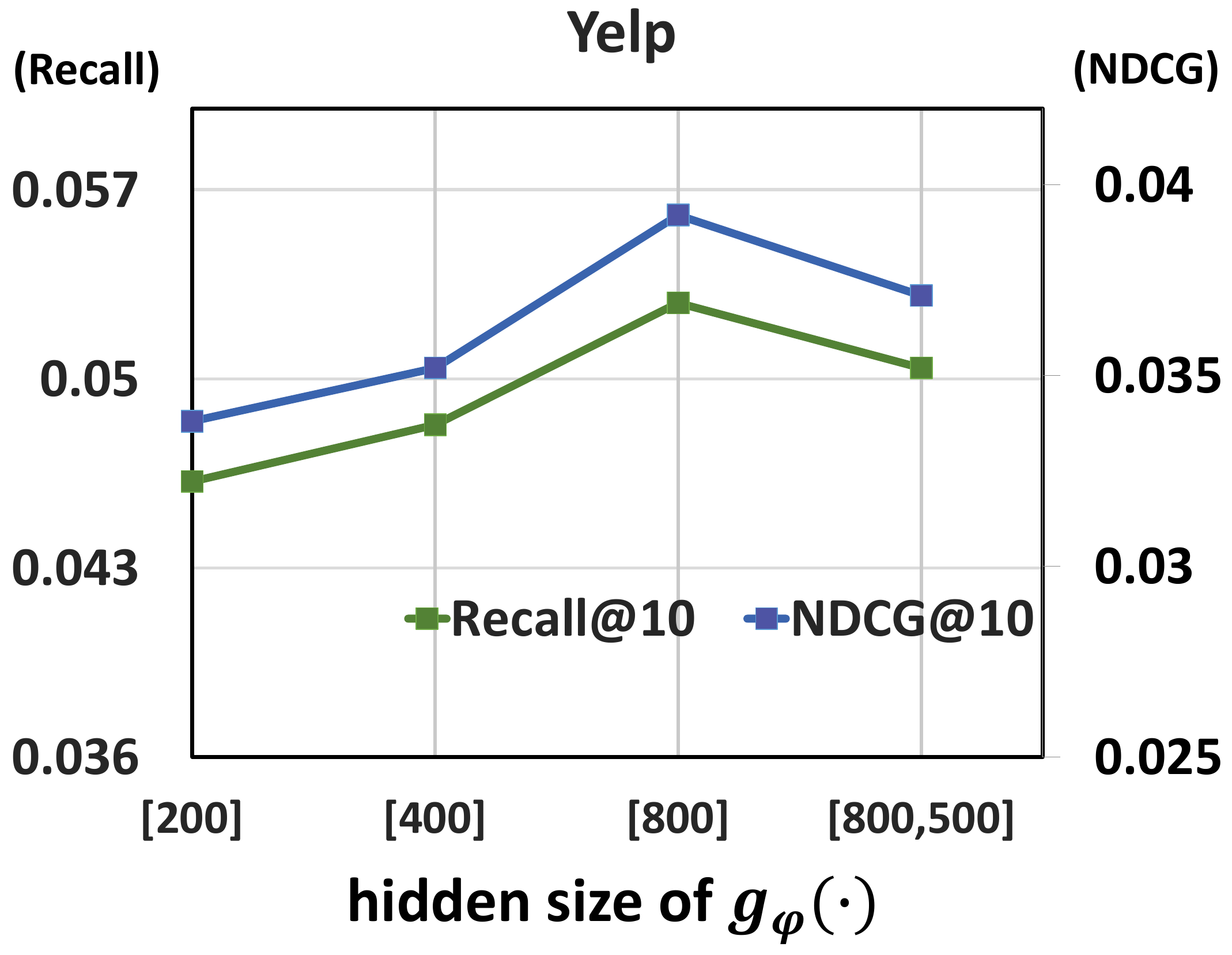}}
  \subfigure{
    \includegraphics[height=1.6in]{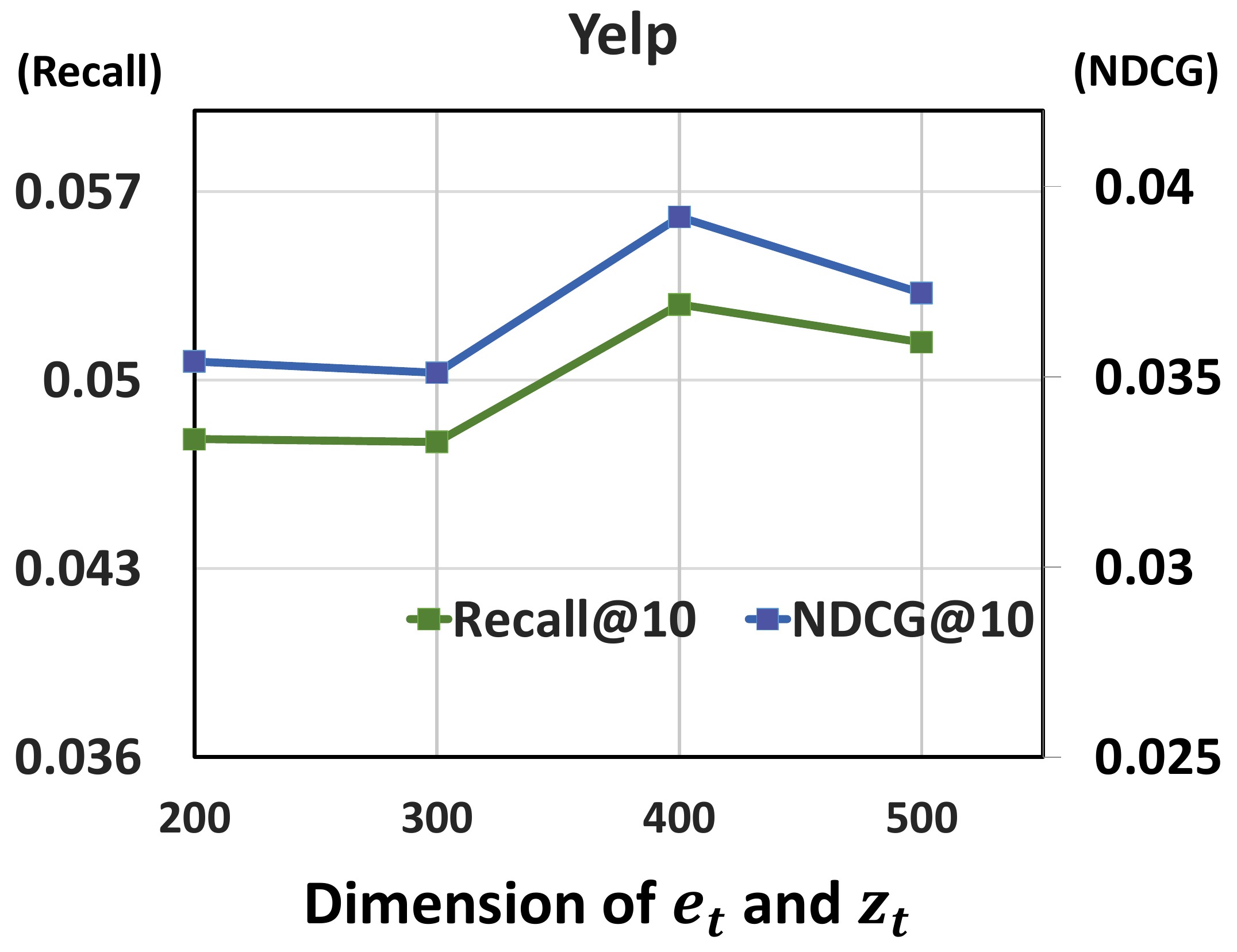}}
  \hspace{-0.7in} 
  \caption{Effect of the hidden sizes of $g_\phi(\cdot)$, $\bm{e}_t$, and $\bm{z}_t$.}
  \label{fig:hidden_size}
\end{figure}

\subsubsection{\textbf{Effect of Hidden Size}}
We conduct experiments to investigate the effect of the hidden sizes of user representations and the latent dimension of VAE networks. In particular, we present the results \wrt varying hidden sizes of $\bm{e}_t$ ($K$), $\bm{z}_t$ ($H$), and the decoder network $g_\phi(\cdot)$.
From the results reported in Figure \ref{fig:hidden_size}, we have the following findings:
\begin{itemize}[leftmargin=*]
    \item When the hidden size of $g_\phi(\cdot)$ changes from $[200]$ and $[400]$ to [800], we can observe that a wider $g_\phi(\cdot)$ yields superior performance. Besides, by comparing the hidden sizes of $\bm{e}_t$ and $\bm{z}_t$ in the range \{200, 300, 400\}, we find a larger size significantly improves Recall@10 and NDCG@10. Such improvements are attributed to enriching the representation abilities of the encoder network and user representations. 

    \item Nevertheless, the model will suffer from over-fitting issue if we blindly increase the number of parameters. For example, the performance drops as we add the layers of $g_\phi(\cdot)$ from $[800]$ to $[800,500]$. 
    Therefore, controlling the parameter number for better OOD generalization is a wise choice. We should carefully tune the hidden sizes to balance the trade-off between the representation ability and over-fitting issue. 
\end{itemize}
We do not show the results on the decoder network because it is implemented by two full-connected layers (\ie $f_{\theta_1}(\cdot)$ and $f_{\lambda}(\cdot)$) whose dimension is decided by $K$, $H$, and the item number $I$ in the dataset. Similar to the encoder network $g_\phi(\cdot)$, we have validated that increasing the layer number of $f_{\theta_1}(\cdot)$ and $f_{\lambda}(\cdot)$ fails to improve the OOD generalization performance.

{\subsection{Case Study (RQ3)}}
\subsubsection{{\textbf{Alignment between User Representation and Shifted Preference}}}
To intuitively understand how CDR captures the preference shifts, we conduct the case study from the user level and population level, respectively. Specifically, for each user, we extract user preference representation $\bm{z}_t$ and the interaction distribution over item categories at each environment. Thereafter, we study whether the shifts of user presentations align with the category-level interaction distribution, which can reflect if the user representations capture preference shifts well.

\begin{figure}[t]
\setlength{\abovecaptionskip}{0.1cm}
\setlength{\belowcaptionskip}{0cm}
\centering 
\includegraphics[width=3in]{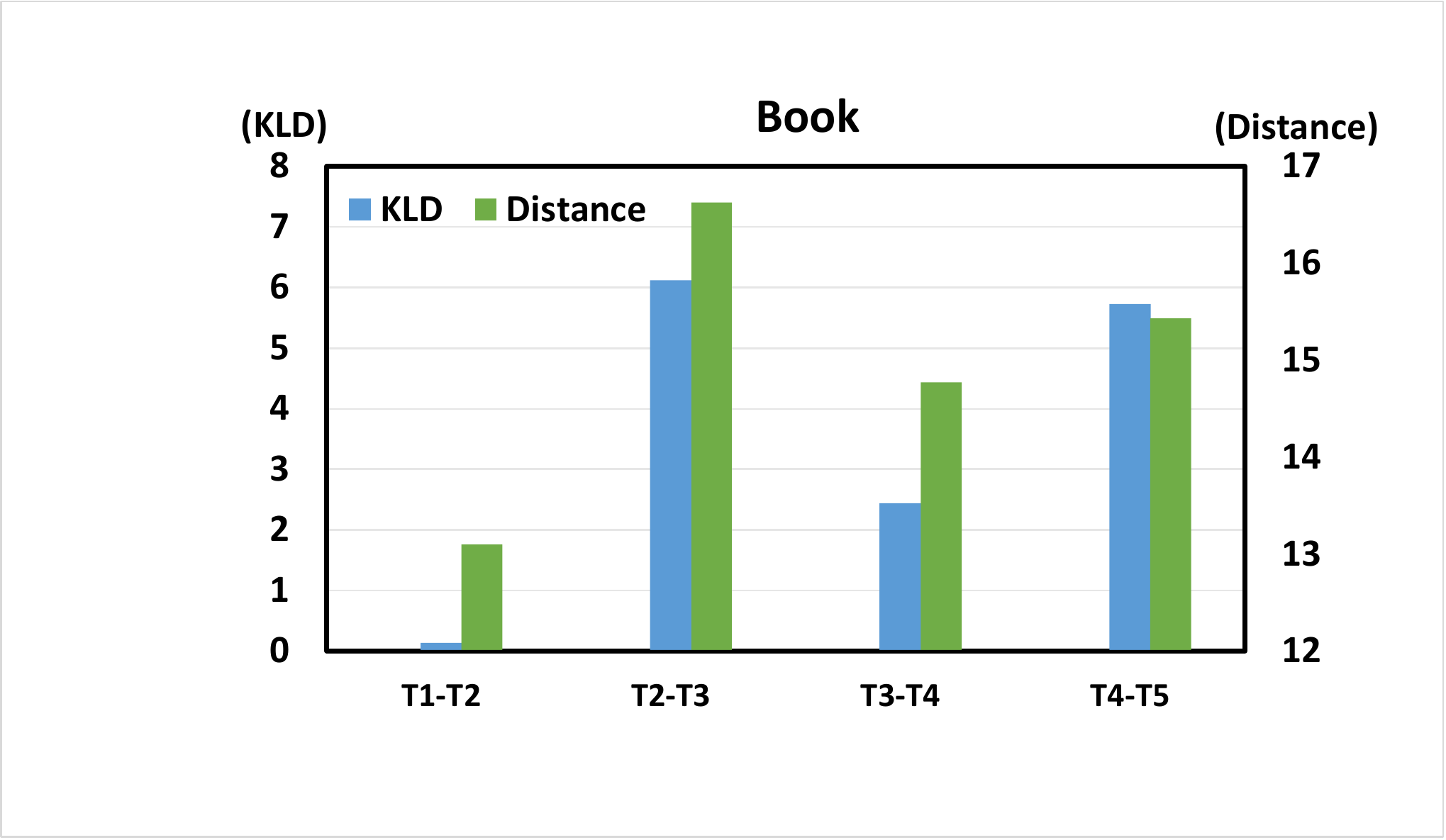}
\caption{Visualization of the KL divergence and Euclidean distance for the sampled user in Book. Note that KL divergence reflects the preference shifts in the interactions while Euclidean distance describes the similarity between the learned user representations.}
\label{fig:case_study_user}
\end{figure}

\begin{figure}[t]
\setlength{\abovecaptionskip}{0.2cm}
\centering
\includegraphics[scale=0.48]{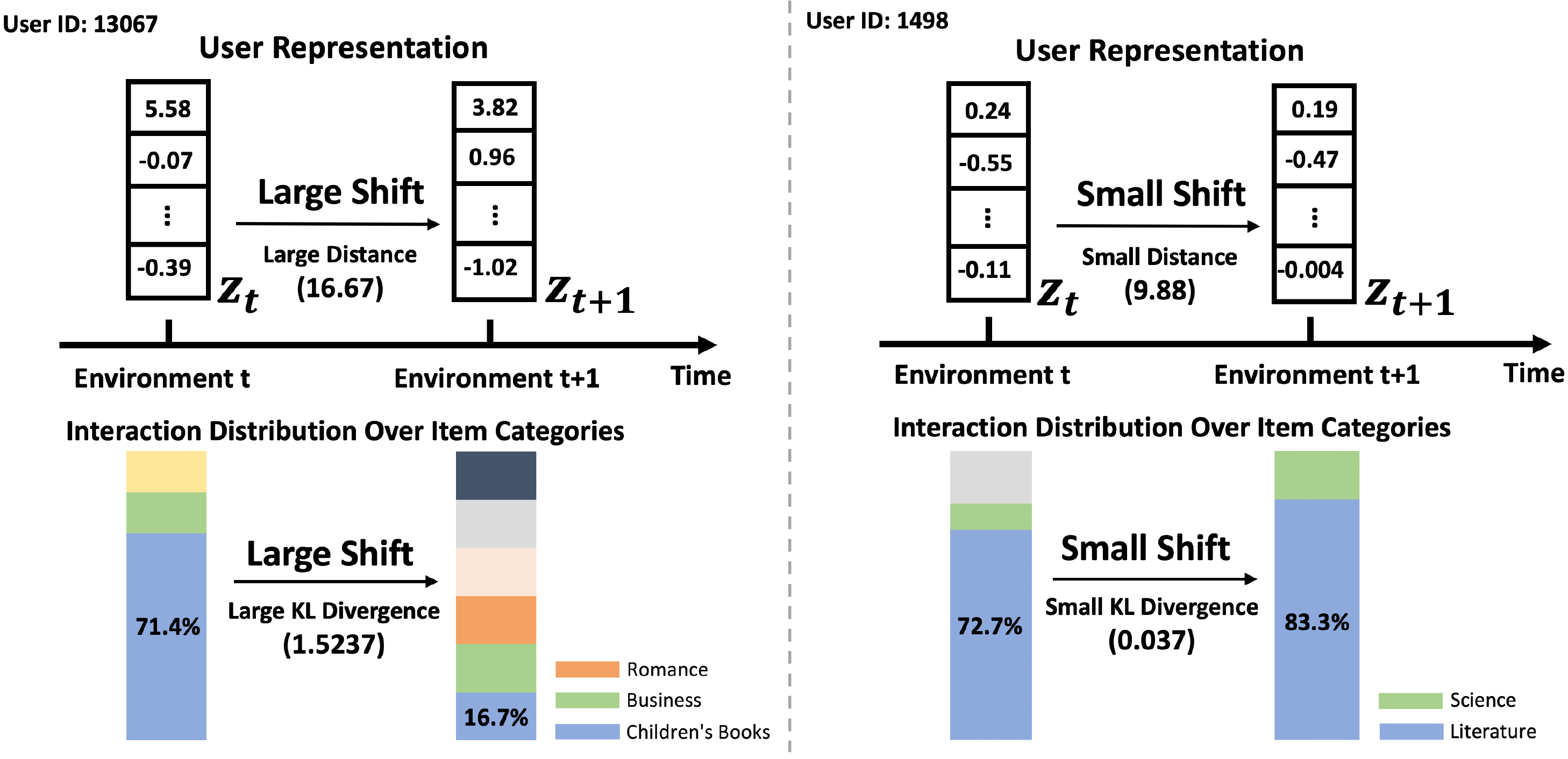}
\caption{Visualization of the interaction distributions and the learned representations for two users in Book. The left figure depicts the user's dramatic shift over category-level preference between two environments. By contrast, the user in the right figure shows relatively stable preference for book genres. The changes of user representations have a good alignment with the user preference shifts.}
\label{fig:case_study_user_2}
\end{figure}

\noindent$\bullet$ \textbf{User-level analysis.} We randomly select a user from the Book dataset who has significant shifts over the category-level interaction distribution.
For the sampled user, we calculate the KL divergence between the category-level interaction distributions in every two continuous environments. Besides, we estimate the user representation similarity between continuous environments by Euclidean distance. The changes of KL divergence and Euclidean distance in five environments are presented in Figure \ref{fig:case_study_user}. 
Furthermore, we also visualize the specific interaction distribution of another two randomly selected users at a more fine-grained level, whose user representations and interaction distributions are provided in Figure \ref{fig:case_study_user_2}.
From the two figures, we have the following conclusions.
\begin{itemize}[leftmargin=*]
    \item The Euclidean distance between user representations has a consistent pattern with the KL divergence between the category-level interaction distributions. For example, when the KL divergence is large in Figure \ref{fig:case_study_user} (\ie T1-T2 and T3-T4), the user representations show the large distance correspondingly. This indicates that user representations are less similar if the preference in the interactions is significantly shifted. In other words, such user representations are capable of capturing the preference shifts.  
    
    \item As to the specific examples in Figure \ref{fig:case_study_user_2}, we can find that the KL divergence well describes the preference shifts over categories: the first user's preference for Children's Books radically drops from 71.4\% to 16.7\% and we have a high KL divergence; by contrast, the interests of the second user are stable, and thus this user has a smaller KL divergence. More importantly, we observe that the user representations have the same distance shifts as the interaction distributions, which is consistent with the findings in Figure \ref{fig:case_study_user}.
\end{itemize}

\begin{figure}[t]
\setlength{\abovecaptionskip}{0.2cm}
\centering
\includegraphics[scale=0.4]{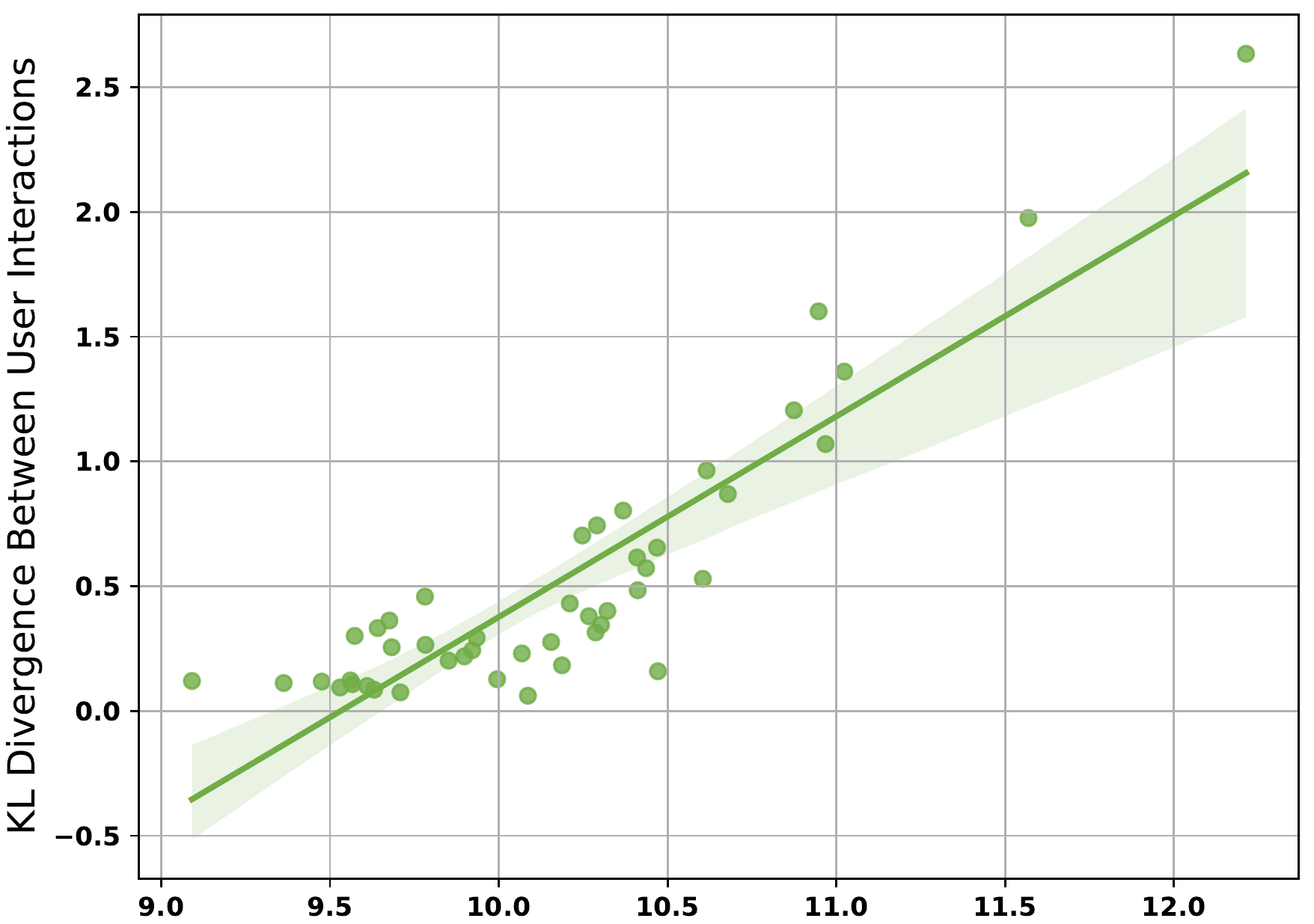}
\caption{Visualization of KL divergence and Euclidean distance at the population level, which shows a strong positive correlation.}
\label{fig:case_study_pop}
\end{figure}

\begin{figure}[t]
\setlength{\abovecaptionskip}{0.2cm}
\centering
\includegraphics[scale=0.48]{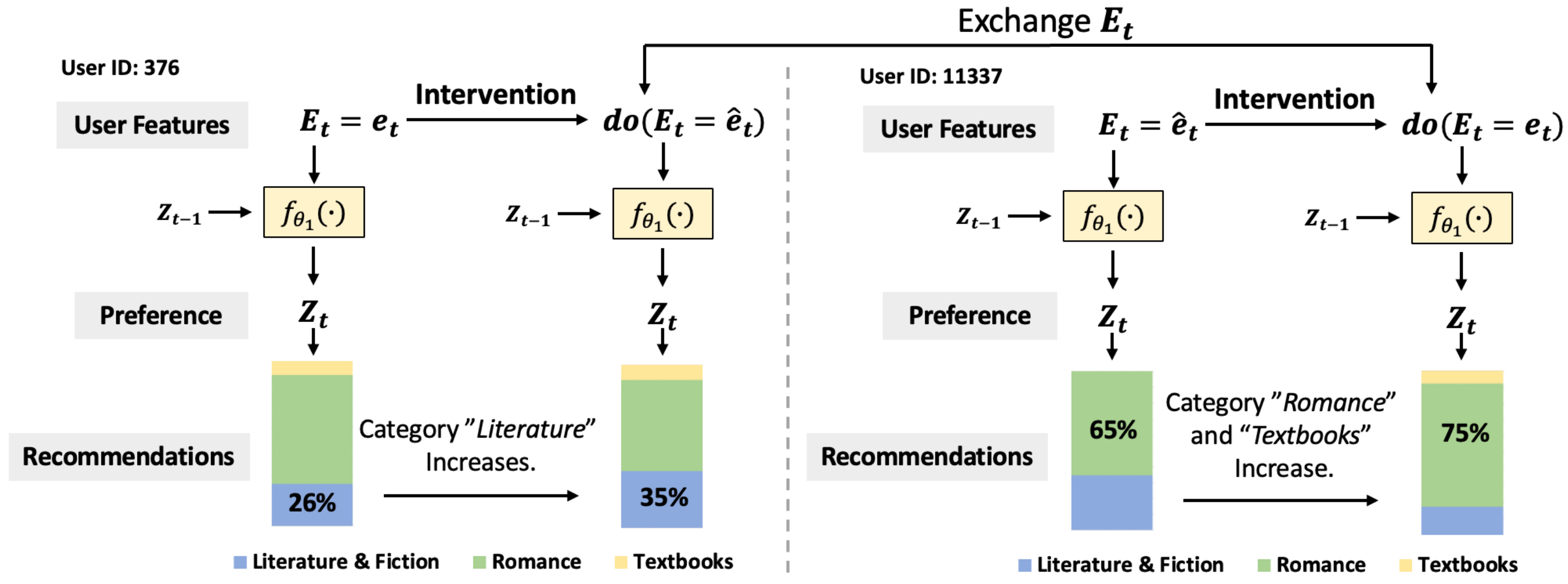}
\caption{{Visualization of the recommendation changes when exchanging $E_t$ of two users in Book. The recommendations are more similar to each other after the intervention.}}
\label{fig:case_intervention}
\end{figure}

\noindent$\bullet$ \textbf{Population-level analysis.} Although the user representations learned by CDR well capture the preference shifts for these users, how does CDR perform over the whole dataset? To answer this question, we do the population-level analysis. 
Specifically, we rank the users in Book according to the average KL divergence in multiple environments, and then divide all users into 50 groups based on the ranking. 
Next, we calculate the average KL divergence and the average Euclidean distance between user representations within each group. Visualization of the correlation between the KL divergence and Euclidean distance is shown in Figure \ref{fig:case_study_pop}, which validates that the euclidean distance between user representations is proportional to the KL divergence between user interactions. This demonstrates that our proposed CDR learns the user representations well to capture the preference shifts in the whole dataset. 

\subsubsection{{{\textbf{Case Study on $do(E_t=\hat{\bm{e}}_t)$}}}}\label{sec:case_do_E}
{We do another case study to inspect whether the do-operation over $E_t$ generates reasonable changes in the recommendations. Specifically, we select two users in the Book dataset, exchange their user features $E_t$, and compare the changes of recommendations. Figure~\ref{fig:case_intervention} shows that their recommendations become similar to each other after the intervention. 
More ``Literature \& Fiction'' books are recommended to User 376 due to the high preference of User 11337 while some textbooks are exposed to User 11337 because User 376 likes them. These observations demonstrate that intervening $E_T$ can affect the predictions of $X_T$ and cause reasonable changes to recommendations in the CDR framework.}

\section{Related Work}
\label{sec:related_work}
In this work, we propose a causal disentangled framework to handle the preference shifts in recommendation, which is closely related to causal recommendation, disentangled recommendation, sequential recommendation, and preference shifts in recommendation. 

\subsection{Causal Recommendation}
In the past decade, data-driven recommender systems have been widely employed to alleviate the issue of information explosion on the Web~[\citealp{wu2019context, causpref, wang2022causal},~{\citealp{saito2022counterfactual}}]. 
Even though great success has been achieved, such data-driven approaches suffer from the issues of bias~[\citealp{zhang2021causal},~{\citealp{wu2022opportunity}}], unfairness~\cite{DiCiccio2020Evaluating}, and filter bubbles~\cite{Ge2020Understanding}.
Recently, the emerging causal approaches shed light on them~[\citealp{,Jadidinejad2021the, zhang2021deep, bonner2018causal,zou2020counterfactual,model-Agnostic},~{\citealp{xu2021causal}}]. 
Specifically, two strands of frameworks receive the most attention: the potential outcome framework~\cite{rubin2005causal} and structural causal models~\cite{pearl2009causality}. 
The potential outcome framework leverages inverse propensity scoring~\cite{saito2020unbiased} or doubly robust~\cite{wang2019doubly} to address the problem of biases in  explicit and implicit feedback~\cite{Zhang2020Large}, such as position bias~\cite{Thorsten2017unbiased} and exposure bias~\cite{saito2020unbiased}. Regarding the structural causal models,  intervention~\cite{zhang2021causal} and counterfactual inference~\cite{wang2021counterfactual,zhang2021cause,zou2020counterfactual} are used to estimate the causal effects~\cite{Pearl2018the} based on the causal relationships. And then the causal effects are more reliable to address recommendation issues, for example, debiasing, unfairness, OOD recommendation, and explanation~\cite{wang2021click, li2021towards, Khanh2021Counterfactual, causpref, wang2022causal}.

Although causal reasoning has been widely applied to recommendation~\cite{xu2023causal,zhu2023causal,gao2022causal,luo2023survey} {and some studies explore the OOD recommendation~\cite{causpref, wang2022causal}, current methods usually utilize user and item features for generalization and ignore the temporal preference shifts across environments.
To fill the gap, this work targets at the under-explored temporal preference shifts without using extra user-item features. }In detail, we discover the generation process of preference shifts under multiple environments and develop a causal framework to alleviate the detrimental effect of preference shifts. 

\subsection{Disentangled Recommendation}
Disentangled recommendation learns the independent representations for the hidden factors (\eg user intention and preference) behind the complex user behaviors, which can bring various merits such as offering explanation or improving the model robustness in the drifted distributions~\cite{ma2020disentangled,Pearl2018the}. 
In order to capture users' diverse preference on items, previous work often disentangles user preference by encouraging the independence of user representations~\cite{ma2020disentangled,wang2020disentangled,wang2020disenhan}. 
For instance, MacridVAE~\cite{ma2019learning} identifies the high-level intention representations for macro disentanglement, and forces each individual dimension in the intention representation to be independent for micro disentanglement. 
Besides, Wang \etal proposed to disentangle the user representations in GCN-based recommender models to model the finer granularity of user intention. 
Lastly, MTIN \cite{jiang2020what} designs a time-aware mask network to distill the interaction sequence and adopts an interest mask network to aggregate fine-grained user preference representations. 

However, previous studies typically learn disentangled representations from the IID data, and thus ignore the significance of capturing robustness across multiple environments~\cite{locatello2020weakly}, decreasing the generalization ability under user preference shifts. 
This work extends the disentangled approaches by considering both the multiple OOD environments and temporal preference shifts across environments. 
Additionally, we reformulate the disentangled representations in recommendation by two matrices and learn the matrices via the sparsity and variance regularization.

\subsection{Sequential Recommendation}
Collaborative filtering-based methods are widely employed in recommender systems~[\citealp{liang2018variational,sarwar2001item},~{\citealp{zhou2019deep,latifi2021session}}], where the user-item matching score is obtained based on user/item representation learned from user historical interactions. However, users' interactions are not independently generated because sequential patterns usually exist within users' consecutive behaviors. Therefore, sequential recommendation, which aims to recommend the next item to a target user, emerges and becomes popular in recent years~\cite{zhu2021learning,manotumruksa2020sequential,quadrana2017personalizing,sachdeva2019sequential}. Early work utilizes Markov Chain to capture the lower-order dependencies~\cite{rendle2010factorizing,he2016fusing}. Later on, deep sequential models (\eg RNN~[\citealp{GRU4Rec},~{\citealp{zhou2018deep}}], CNN~\cite{Caser}, Transformer~\cite{SASRec,Bert4Rec}, GNN~\cite{SRGNN,SURGE} and others~\cite{sabour2017dynamic}) are employed to capture the higher-order dependencies. In addition to general sequential methods, CauseRec~\cite{zhang2021cause} conditionally constructs the counterfactual interaction sequences, and then performs contrastive representation learning by using both counterfactual and observational data. 
DSSRec~\cite{ma2020disentangled} disentangles the intentions behind the user interaction sequence, and constructs seq2seq training samples by using only pairs of sub-sequences with the same intention, leading to better sequential modeling. 
Recently, ACVAE \cite{xie2021adversarial} has introduced adversarial learning to the variational Bayes framework for sequential recommendation and utilized contrastive learning to learn better user representations.

Different from sequential recommendation, CDR focuses on the generalizable recommendation to handle preference shifts, which aims to predict the user preference in the new OOD environment instead of next-item recommendation. Technically, CDR learns the invariant user preference in a short period while capturing temporal preference shifts between multiple environments. Furthermore, CDR formulates a sparse structure from the preference representation to interaction prediction for more robust disentanglement.

\subsection{Preference Shifts in Recommendation}
User preference may shift over time for many reasons, including the changes of user features (\eg income and pregnancy) and environment factors (\eg seasonal variation). 
For example, in the scenario of food recommendation, a user might become liking expensive but healthy food if the user's income increases. 
Since such preference shifts are frequent in the real-world scenarios, the recommender models should update the user representations adaptively over time. 
Ignoring the shifting nature of preference to learn user representation will lead to inappropriate recommendations. In addition to disentangled recommendation and sequential recommendation, there exists some work with the potential of addressing this issue. 
In particular, Aspect-MF \cite{zafari2019modelling} analyses the dynamicity of temporal preference aspects using a component-based approach, and identifies the aspects that are easy to drift. 
ST-LDA \cite{yin2016adapting} learns region-dependent personal interests and crowd preference to adapt to preference shifts. 
Lastly, MTUPD \cite{wangwatcharakul2021novel} utilizes a forgetting curve function to calculate the correlations of user preference in different time periods.

\vspace{5pt}
\noindent$\bullet\quad$\textbf{Long- and short-term interest modeling.}
Another possibly related direction considers the modeling of both long- and short-term interests, where short-term interests are inferred from recent interactions to capture the preference shifts, and long-term interests represent the stable preference over time.
Many studies have shed light on the modeling of long- and short-term user interests, for instance, PLSPL \cite{sritrakool2021personalized} utilizes attention mechanism to characterize the long-term preference while integrating the location and category information to capture the short-term preference. 
LSTPM \cite{sun2020go} develops a context-aware nonlocal network structure to explore the temporal and spatial correlations in users' trajectories for the long-term preference, and adopts a geo-dilated RNN to fully exploit the geographical relations for the short-term preference. 
Lastly, KERL \cite{wang2020kerl} fuses knowledge graph into a reinforcement learning framework to capture long-term preference and predict short-term interests. 
Although the short-term preference might infer the preference shifts, such work has not discovered the causal reasons for invariant and shifted preference. By contrast, we model the invariant and shifted preference simultaneously by leveraging their underlying causal relations and conduct the disentangled preference learning from multiple OOD environments.


\vspace{5pt}
\noindent$\bullet\quad$\textbf{Domain adaptation.}
Domain adaptation has been widely applied to solve the problem of distribution shifts, which improves the adaptation ability by using less data.
In recommendation, its main application scenarios include cross-domain recommendation \cite{zhu2021personalized, zhao2019cross, zhao2020catn} and cold-start recommendation \cite{yuan2020parameter}. 
Technically, model adaptation can be implemented by parameter patch \cite{sheng2021one, yuan2020parameter}, feature transformation \cite{lin2021task}, and meta learning \cite{yu2021personalized}.
These techniques have been well studied to estimate the preference of the users in a new domain or the cold-start users within the same domain. However, the preference shift in OOD recommendation is totally different because it is related to the same users with dynamic interests over time. This requires us to consider the connection between the interactions in different environments, \ie the cross-environment preference learning. 


\section{Conclusion and Future Work}
\label{sec:conclusion}

In this work, we formulated the preference shifts from a causal view and inspected the underlying causal relations from the perspective of multiple environments. Based on the causal relations, we proposed the CDR framework, which captures the preference shifts across environments via a temporal VAE and learns the sparse structure from user preference to interactions. In particular, CDR leverages two learnable matrices to disentangle user representations and formulate the sparse structure. We optimized CDR by a multi-objective loss with variance and sparsity regularization. 
During the training and inference stages, we could flexibly adjust the number of environments to balance the learning of shifted and invariant preference on different datasets. Extensive experiments validate the effectiveness of CDR in capturing preference shifts and achieving superior generalization performance than the baselines. Furthermore, the in-depth analysis demonstrates the rationality of using multiple environments, the sparse structure, and the multi-objective loss.

This work attempts to learn user preference from multiple environments for handling user preference shifts. In this light, there are many promising directions in future work. In particular, 1) as discussed in Section~\ref{sec:environment}, we equally split user interactions into $T$ environments to simplify the data pre-processing. Nevertheless, it can be improved by developing more effective but complex methods to divide environments, \eg clustering interactions according to the time interval. 
2) It is meaningful to discover more fine-grained causal relations in Figure \ref{fig:causal_graph}, such as the mutual impact between the category-level user preference. Besides, we might utilize observed user features (\eg age) to study more fine-grained relations between $E_t$ and $Z_t$. And 3) it is non-trivial to incorporate item features for the disentangled representation learning, which may help CDR to distinguish the category-level user preference.




{
\tiny
\bibliographystyle{ACM-Reference-Format}
\balance
\bibliography{bibtex}
}
\appendix

\end{document}